\documentstyle[12pt,epsfig]{article}
\def\bge{\begin{equation}}
\def\ene{\end{equation}}
\def\bg{\begin{eqnarray}}
\def\en{\end{eqnarray}}
\def\nn{\nonumber}
\def\ra{\rightarrow}

\begin{document}
\renewcommand{\thefootnote}{\fnsymbol{footnote}} 
\begin{flushright}
ADP-96-17/T220
\end{flushright}
\vspace{0.5cm}
\begin{LARGE}
\begin{center}
Self-consistent description of finite nuclei based on a 
relativistic quark model
\end{center}
\end{LARGE}
\vspace{2cm}
\begin{center}
\begin{large}
Koichi~Saito\footnote{ksaito@nucl.phys.tohoku.ac.jp} \\
{\sl Physics Division, Tohoku College of Pharmacy \\
Sendai 981, Japan} \\
\vspace{0.5cm}

Kazuo~Tsushima\footnote{ktsushim@physics.adelaide.edu.au} and 
Anthony~W.~Thomas~\footnote{athomas@physics.adelaide.edu.au} \\
{\sl Department of Physics and Mathematical Physics, \\
 and Institute for Theoretical Physics, \\
University of Adelaide, South Australia 5005, Australia}
\end{large}
\end{center}
\vspace{4.0cm}
\newpage
\begin{abstract}
Relativistic Hartree equations for spherical nuclei have been derived 
from a 
relativistic quark model of the structure of bound nucleons which 
interact through the (self-consistent) exchange of scalar ($\sigma$) and 
vector ($\omega$ and $\rho$) mesons. 
The coupling constants and the mass of the $\sigma$-meson are determined 
from the properties of 
symmetric nuclear matter and the rms charge radius in $^{40}$Ca.  
Calculated properties of static, closed-shell nuclei from $^{16}$O to 
$^{208}$Pb are compared with experimental data and with results of Quantum 
Hadrodynamics (QHD).  The dependence of the results 
on the nucleon size and the quark mass is 
investigated.  Several possible 
extensions of the model are also discussed. 
\end{abstract}
\newpage
\section{Introduction}
\label{Intro}
{\em Do quarks play an important role in finite nuclei ?}  
This is one of the central questions 
in nuclear physics.  We now know that explicit quark 
degrees of freedom are certainly needed to understand deep-inelastic 
scattering at momentum transfers of several GeV, e.g., the nuclear EMC 
effect\cite{emc}. However, it is perhaps not surprising that such 
physics seems irrelevant at a scale of a few tens of MeV, and the idea 
that the quark structure of the nucleon should be of any importance in 
nuclear physics is often dismissed out of hand. 

Within the traditional approach to nuclear physics 
these tens of MeV are the result of a fairly fine-tuned cancellation 
between short-range repulsion, fairly strong intermediate range 
attraction and 
large kinetic energy. Nearest 
neighbour nucleons in nuclear matter tend to be an average 1.8 fm apart, 
which corresponds to the intermediate range attraction associated with 
two-pion-exchange. This attraction is mainly generated by the coupling to 
the $\Delta$ (through the process NN $\ra$ N$\Delta \ra$ NN), 
which is just the first excited state of the internal quark structure of 
the nucleon, namely spin-flip.  Of course, one can treat the 
NN $\ra$ N$\Delta$ system in a coupled channels 
formalism without ever mentioning quark structure. One can even replace the 
N$\Delta$ coupling by simple scalar meson ($\sigma$) exchange without 
significantly altering the binding energy of nuclear systems. If this were
the full story quarks would be irrelevant.  

However, it is well known that two-body interactions alone cannot reproduce 
either the saturation properties of nuclear matter or the binding energies 
and charge radii of finite nuclei. One needs some form of three-body force, 
and this leads to a nightmare of badly known form-factors and coupling 
constants, of inconsistent treatments of the $\pi$N, NN and NNN 
systems, and so on\cite{formf}. 

Conventional descriptions of finite nuclei are usually based on the 
Schr{\"o}dinger equation which involves nucleons interacting through 
static potentials.  
These effective potentials are sometimes determined partly by the 
Brueckner-Hartree-Fock equations, and often involve some phenomenological 
dependence on the nuclear density to include the effect of higher-order 
contributions to the N-N force in nuclei\cite{nonrel}.  Those 
density-dependent 
potentials are however {\em purely phenomenological}.  Although those 
conventional approaches give rather successful results for the ground 
states of nuclei, discrepancies between experimental data and theoretical 
calculations still remain.  

Furthermore, there are several problems at a 
low energy scale, which might not be explained by the traditional approach. 
One example is the Okamoto-Nolen-Schiffer (ONS) anomaly\cite{nsa}, 
which is a well known, long-standing problem in nuclear physics.  
Conventional nuclear contributions to the anomaly are thought to be at 
the few percent level and cannot explain the experimental 
findings\cite{nsa2}.  The effects of charge symmetry breaking in 
the nuclear force, especially $\rho$-$\omega$ mixing, seem to reduce the 
discrepancy\cite{nsa,nsa2,hs}.  However, recent investigations of 
the off-shell variation of the $\rho$-$\omega$ mixing amplitude have put 
this explanation into question\cite{ght}.  

Another example concerns the unitarity of the Cabibbo-Kobayashi-Maskawa 
(CKM) matrix elements.  
There is also a well known discrepancy between the value of the 
Fermi decay constant extracted from super-allowed Fermi beta-decay of nuclear
isotriplets and that required by unitarity of the 
CKM matrix. This discrepancy remains at the level
of a few tenths of a percent (but beyond two standard deviations from the 
experimental data) after the most rigorous investigation of 
conventional nuclear and radiative corrections\cite{weak}. 

These facts seem to indicate that the traditional 
approach may have its limitations and suggest a need for 
the study of alternate approaches including 
sub-nucleonic degrees of freedom.  

Exciting a single quark in the nucleon costs about 400 MeV. This is not 
significantly different from the energy required to excite a $\Delta$.
It is the same order of magnitude as the scalar and vector potentials 
required in Quantum Hadrodynamics (QHD)\cite{qhd}.  
Furthermore, the quarks are very light and should be able to 
respond rapidly to their environment. 
{\em We know of no physical argument why this response should be ignored.} 

About eight years ago Guichon\cite{guichon} proposed an entirely different 
model for nuclear matter, based on a mean-field description of nucleons 
described by the non-overlapping MIT bag model, namely the quark-meson 
coupling (QMC) model.  Later this model was refined by Saito and 
Thomas\cite{st1} who clarified its relationship to QHD. 
Related work has been carried out by a number of groups\cite{yazaki}. 

The QMC model may be viewed as an extension of QHD in which the 
nucleons still interact through the exchange of $\sigma$ and $\omega$ 
mesons.  However, the mesons couple not to point-like nucleons but to
confined quarks.  In studies of infinite nuclear matter\cite{guichon,st1} 
it was found that
the extra degrees of freedom provided by the internal structure of the
nucleon mean that one gets quite an acceptable value for the
incompressibility once the coupling constants are chosen to
reproduce the correct saturation energy and density for symmetric 
nuclear matter. This is a significant improvement on QHD 
at the same level of sophistication. 
A wonderful feature of this picture is that most specific details 
of the model are irrelevant. For example, it does not matter that one 
uses the MIT bag model, all that is needed is that the quarks are 
relativistic. It does not matter that the $\omega$ generates the 
repulsion, all that is needed is that it has a vector character.

There have been several interesting applications to the properties of 
finite nuclei using the local-density approximation.  
Surprisingly, the model can provide a semi-quantitative explanation 
of the ONS anomaly when quark mass differences are 
included\cite{st2,juelich}. An application of the model, including quark mass 
differences, has also suggested a previously unknown correction to the 
extraction of the CKM matrix element, $V_{ud}$, from super-allowed Fermi 
beta-decay, which would bring the discrepancy in the unitarity problem of the 
CKM matrix back to only one standard deviation\cite{st3,osaka}. 
Furthermore, in the light of current experimental work in relativistic 
heavy ion collisions\cite{brwn,expts}, 
which produce nuclear matter at densities several times 
normal, there has been some initial work on the variation of baryon and 
meson properties with density\cite{hadrons}. 
Finally the model has been applied to the case where quark degrees of freedom 
are undisputedly involved -- namely the nuclear EMC effect\cite{st4}. 

However, the inherent problems of the local-density approximation mean
that these applications can at best be semi-quantitative and it was
clearly very important that the extension to finite nuclei be developed. 
Recently we have succeeded in developing a formulation 
of the quark-meson coupling 
model for finite nuclei\cite{finite}, based on the Born-Oppenheimer
approximation.  The effective equation of motion for the nucleon, as 
well as self-consistent equations for the meson mean fields have been 
derived.  We have also shown some initial results for $^{16}$O and 
$^{40}$Ca, and discussed, in particular, the spin-orbit force in the 
model and its relation to the corresponding force in conventional models 
involving meson exchange between point-like nucleons. 
Related work has been carried out by several groups\cite{jin,blun}. 

Our aim in this paper is to 
show details of our 
calculations for the properties of spherically symmetric, closed-shell 
nuclei from $^{16}$O to $^{208}$Pb.  

In Sect.\ref{qmc} we briefly summarize
the QMC model for both infinite nuclear 
matter and finite nuclei.  Starting from the Lagrangian density proposed in 
Ref.\cite{finite} we derive the Dirac equation for the nucleon and the 
equations for the meson fields in mean-field approximation and discuss 
the effective $\sigma$-N coupling constant.  The properties of infinite 
nuclear matter are then reviewed briefly in Sect.\ref{matter}.  
We introduce the 
couplings of the $\rho$ meson and the photon to the quarks,
in Sect.\ref{fnt}, in order to describe 
realistic finite nuclei, and again derive explicit forms for the equations 
of motion of the nucleon and the mesons.  In Sect.\ref{results}, the
parameters in our calculations are determined so as to reproduce the rms 
charge radius of $^{40}$Ca, and then our calculated results are 
compared with those of QHD and the experimental data.   
Our results agree with 
data favorably.  We give a summary and discuss some remaining problems 
and some possible extensions in Sect.\ref{discuss}.  

\section{Relativistic formulation of the QMC model}
\label{qmc}
The solution of the general problem of a composite, quantum particle 
moving in background scalar and vector fields that vary with position is 
extremely difficult.  One has a chance to solve the particular problem of 
interest to us, namely light quarks confined in a nucleon which is 
itself bound in a finite nucleus, only because the nucleon motion is 
relatively slow and the quarks highly relativistic. Thus the 
Born-Oppenheimer approximation, in which the nucleon internal 
structure has time to adjust to the local fields, is naturally suited to 
the problem. It is relatively easy to establish that the method should 
be reliable at the level of a few percent\cite{finite}. 

Even within the Born-Oppenheimer approximation, the nuclear surface 
gives rise to external fields that may vary appreciably 
across the finite size of the nucleon.  In Ref.\cite{finite}, our 
approach was to start with a classical nucleon 
and to allow its internal structure (quark wavefunctions and bag radius) 
to adjust to minimise the energy of three quarks in the ground-state of 
a system consisting of the bag plus constant scalar and vector fields, 
with values equal to those at the centre of the nucleon. 
Of course, the major problem with 
the MIT bag (as with many other relativistic models of nucleon structure) 
is that it is difficult to boost. We therefore solve the bag 
equations in the instantaneous rest frame (IRF) of the nucleon -- using a
standard Lorentz transformation to find the energy and momentum of the
classical nucleon bag in the nuclear rest frame. 
Having solved the problem using the meson fields at the centre of 
the $\lq\lq$nucleon'' (which is a quasi-particle with nucleon quantum 
numbers), 
one can use perturbation theory to correct for the variation of the 
scalar and vector fields across the nucleon bag. In first order perturbation 
theory only the spatial components of the vector potential 
give a non-vanishing contribution. (Note that, although in the nuclear 
rest frame only the time component of the vector field is non-zero, 
in the IRF of the nucleon there are also non-vanishing spatial components.) 
This extra term is a correction to the spin-orbit force, which is 
however very 
small\cite{finite}.  Blunden and Miller\cite{blun} have also
investigated 
the QMC model for finite nuclei recently.
They too reported 
that the correction due to the variation of meson fields inside the 
nucleon is not large.  

As shown in Ref.\cite{finite}, the basic result in the QMC model is that, 
in the scalar ($\sigma$) and vector ($\omega$) 
meson fields, the nucleon behaves essentially as a point-like 
particle with an effective mass 
$M_N^{\star}$, which depends on the position through only the $\sigma$ 
field, moving in a vector potential generated by the $\omega$ meson. 
Because of the vector character, the vector interactions have {\em no effect 
on the nucleon structure} except for an overall phase in the wave function, 
which gives a shift in the nucleon energy.  

If we restrict ourselves to consider static, spherically 
symmetric nuclei, the effective Lagrangian density involving the quark 
degrees of freedom and the meson fields in mean-field approximation would 
be given by\cite{finite}
\bg
{\cal L}_{QMC}^0 &=& \overline{\psi} [i \gamma \cdot \partial 
- M_N^{\star}(\sigma({\vec r})) - g_\omega \omega({\vec r}) \gamma_0] 
\psi \nn \\
&-& \frac{1}{2}[ (\nabla \sigma({\vec r}))^2 + 
m_{\sigma}^2 \sigma({\vec r})^2 ] 
+ \frac{1}{2}[ (\nabla \omega({\vec r}))^2 + m_{\omega}^2 
\omega({\vec r})^2 ] ,
\label{relat}
\en
where $\psi({\vec r})$, $\sigma({\vec r})$ and $\omega({\vec r})$ are 
respectively the 
nucleon, $\sigma$ and the time component of $\omega$ fields. (In the mean 
field approximation, the meson fields are given by their time independent 
expectation values in the ground state of the nucleus.  For a symmetric, 
parity-eigenstate, the spatial component of the $\omega$ field vanishes.)  
$m_\sigma$ and $m_\omega$ are respectively the masses of the 
$\sigma$ and $\omega$ 
mesons. $g_\omega$ is the $\omega$-N coupling constant (which is related to 
the corresponding quark-$\omega$ coupling constant, 
$g_\omega^q$, as $g_\omega = 
3 g_\omega^q$\cite{guichon,st1,finite}).

The effective nucleon mass $M_N^{\star}$ is given by a model describing the 
nucleon structure and it depends on the position through only the $\sigma$ 
field (as mentioned above).  We use the MIT bag model in this paper.  
(A relativistic oscillator model has also been used in Ref.\cite{blun} as an 
alternative model.)  If we define the field-dependent $\sigma$-N coupling 
constant, $g_\sigma(\sigma)$, by
\bge
M_N^{\star}(\sigma({\vec r})) \equiv M_N - g_\sigma(\sigma({\vec r})) 
\sigma({\vec r}) , \label{coup}
\ene
where $M_N$ is the free nucleon mass, it is easy to compare with 
QHD\cite{qhd}.  The Lagrangian density then becomes
\bg
{\cal L}_{QMC}^0 &=& \overline{\psi} [i \gamma \cdot \partial 
- M_N + g_\sigma (\sigma({\vec r})) \sigma({\vec r}) 
- g_\omega \omega({\vec r}) \gamma_0] \psi \nn \\
&-& \frac{1}{2}[ (\nabla \sigma({\vec r}))^2 + 
m_{\sigma}^2 \sigma({\vec r})^2 ] 
+ \frac{1}{2}[ (\nabla \omega({\vec r}))^2 + m_{\omega}^2 
\omega({\vec r})^2 ] . 
\label{relat2}
\en
The difference from QHD clearly lies in the fact that the internal 
structure of the nucleon has forced a {\em known} dependence of the 
$\sigma$-N coupling constant on the scalar field itself.  (Note that this 
dependence is not the same as that adopted in the density-dependent 
hadron field theory\cite{fuchs} where the meson-nucleon vertices are 
assumed to depend directly on the baryonic densities.)  

Equation (\ref{relat}) yields the Dirac equation for the nucleon
\bge
[i\gamma \cdot \partial -M_N^{\star}(\sigma)- 
g_\omega \gamma_0 \omega ] \psi = 0 ,
\label{dirac}
\ene
as well as the equations for the meson mean-fields
\bg
(-\nabla^2_r+m^2_\sigma)\sigma(\vec{r})&=& - \left( 
\frac{\partial}{\partial \sigma}M_N^{\star}(\sigma) \right) 
\rho_s({\vec r}), \label{sig} \\
(-\nabla^2_r+m^2_\omega) \omega(\vec{r}) &=&  
g_\omega \rho_B({\vec r}), \label{omg}
\en
where $\rho_s$ and $\rho_B$ are respectively the scalar and baryon densities 
of the nucleon in the nucleus, $A$, which are defined by the expectation 
values in the ground state of $A$: 
\bge
\rho_s({\vec r}) = \langle A|\overline{\psi}\psi(\vec{r})|A \rangle \ \ 
\mbox{and} \ \ \rho_B({\vec r}) = \langle A|\psi^\dagger\psi(\vec{r})|A 
\rangle . \label{density}
\ene

One can easily see that the derivative of $M_N^{\star}$ with respect to 
$\sigma$ in Eq.(\ref{sig}) is the response of the nucleon to the 
external scalar field, and that it is given by the scalar density of a 
quark in the nucleon\cite{st1}: 
\bge
\left( \frac{\partial M_N^{\star}}{\partial \sigma} \right) 
= -3g_{\sigma}^q \int_{bag} d{\vec r} \ {\overline \psi}_q \psi_q 
\equiv -3g_{\sigma}^q S(\sigma) , \label{deriv}
\ene
where $g_\sigma^q$ is the quark-$\sigma$ coupling constant and 
$\psi_q$ is the quark wave function in the nucleon bag with 
radius $R_B^{\star}$ in matter (we denote the bag radius of the free nucleon 
by $R_B$).  We define $S(\sigma)$ by Eq.(\ref{deriv}), 
which is a function of scalar field itself. 
Using the MIT bag model\cite{finite}, $S$ is explicitly given by 
\bge
S(\sigma(\vec{r})) = \frac{\Omega/2 + m_q^{\star}R_B^{\star}(\Omega-1)}
{\Omega(\Omega-1) + m_q^{\star}R_B^{\star}/2}, \label{sss}
\ene
where $\Omega = \sqrt{x^2 + (R_B^{\star}m_q^{\star})^2}$ is the kinetic 
energy of 
the quark and $m_q^{\star}({\vec r})$ is the effective quark mass defined 
by $m_q - g_\sigma^q \sigma({\vec r})$, with the quark mass $m_q$ 
inside the free nucleon.  
Furthermore, we define the scalar density ratio, $S(\sigma)/S(0)$, 
by $C(\sigma)$ and the $\sigma$-N coupling constant at $\sigma=0$ by 
$g_\sigma$ (i.e., $g_\sigma \equiv g_\sigma(\sigma=0)$)\footnote{
Note the change in notation from the earlier works of Saito and Thomas where 
$C$ was used for what we now call $S$. }:
\bge
C(\sigma) = S(\sigma)/S(0) \ \ \mbox{and} \ \ 
g_{\sigma} = 3g_{\sigma}^q S(0) . \label{cn}
\ene
Comparing with Eq.(\ref{coup}), we find 
\bge
\left( \frac{\partial M_N^{\star}}{\partial \sigma} \right) 
= -g_{\sigma} C(\sigma) = - \frac{\partial}{\partial \sigma}
\left[ g_\sigma(\sigma) \sigma \right],
\label{deriv2}
\ene
and the equation for the $\sigma$ field becomes
\bge
(-\nabla^2_r+m^2_\sigma)\sigma(\vec{r}) = g_\sigma C(\sigma) 
\rho_s({\vec r}) . \label{sig2} 
\ene
We will discuss the $\sigma$-N coupling constant further in the next 
subsection. 

\subsection{Nuclear matter limit}
\label{matter}
In isospin-symmetric, infinite nuclear matter the sources of the fields 
are constant and can be related to the nucleon Fermi momentum 
$k_F$\cite{qhd}: 
\bg
\rho_B &=& \frac{4}{(2\pi)^3}\int d\vec{k} \theta (k_F - k) = 
\frac{2 k_F^3}{3\pi^2} , 
\label{rhoB}\\
\rho_s &=& \frac{4}{(2\pi)^3}\int d\vec{k} \theta (k_F - k)
\frac{M_N^{\star}}{\sqrt{M_N^{\star 2}+\vec{k}^2}},
\label{rhos}
\en
where $M_N^{\star}$ means the constant value of the effective nucleon 
mass given by the MIT bag model (detailed derivations are given in 
Ref.\cite{finite}). 

As in Ref.\cite{finite}, the bag constant $B$ and the 
parameter $z_0$ (which accounts for the sum of the c.m. and 
gluon fluctuation corrections) in the familiar form of the MIT bag 
model Lagrangian are 
fixed to reproduce the free nucleon mass ($M_N$ = 939 MeV) 
under the stationary 
condition in terms of the free bag radius, $R_B$.  In the following we 
treat the free bag radius as a variable parameter to test 
the sensitivity of our 
results to the free nucleon size.  The results for $B$ and $z_0$ are 
shown in Table~\ref{b,z}.  We choose the free quark mass $m_q$ = 0, 5, 
10 MeV. 
\begin{table}[htbp]
\begin{center}
\caption{Bag constant and parameter $z_0$ for several bag radii and quark 
masses.}
\label{b,z}
\begin{tabular}[t]{c|ccc|ccc|ccc}
\hline
$m_q$(MeV) & & 0 & & & 5 & & & 10 & \\
\hline
$R_B$(fm) & 0.6 & 0.8 & 1.0 & 0.6 & 0.8 & 1.0 & 0.6 & 0.8 & 1.0 \\
\hline
$B^{1/4}$(MeV) &211.3&170.3&144.1&210.9&170.0&143.8&210.5&169.6&143.5 \\
$z_0$     &3.987&3.273&2.559&4.003&3.295&2.587&4.020&3.317&2.614 \\
\hline
\end{tabular}
\end{center}
\end{table}

Let ($\overline{\sigma}, \overline{\omega}$) be the {\em constant} 
mean-values of the meson fields. From Eqs.(\ref{omg}) and (\ref{sig2}) 
we find
\bg
\overline{\omega}&=&\frac{g_\omega \rho_B}{m_\omega^2} , \label{omgf}\\
\overline{\sigma}&=&\frac{g_\sigma }{m_\sigma^2}C(\overline{\sigma})
\frac{4}{(2\pi)^3}\int d\vec{k} \theta (k_F - k) 
\frac{M_N^{\star}}{\sqrt{M_N^{\star 2}+\vec{k}^2}} , \label{sigf}
\en
where $C(\overline{\sigma})$ is now the constant value of $C$  in the 
scalar field.  
This self-consistency equation for $\overline{\sigma}$ 
is the same as that in QHD, {\it except} that in the latter model one has 
$C(\overline{\sigma})=1$ \cite{st1,blun}. (This corresponds to the
heavy-quark mass limit of the QMC model.)
%
%
\begin{figure}[htb]
\centering{\
\epsfig{file=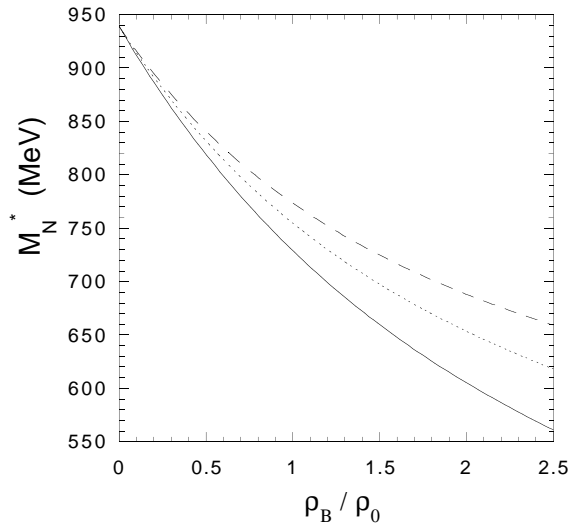,height=7cm}
\caption{Effective nucleon mass as a function of density (for $m_q$=5MeV). 
The solid, dotted and dashed curves correspond to
$R_B$ = 0.6, 0.8 and 1.0 fm, respectively.}
\label{fig:mass}}
\end{figure}
%
%

Once the self-consistency equation for $\overline{\sigma}$ has been solved, 
one can evaluate the total energy per nucleon\cite{guichon,st1,finite}: 
\bge
E^{total}/A=\frac{4}{(2\pi)^3 \rho_B}\int d\vec{k} 
\theta (k_F - k) \sqrt{M_N^{\star 2}+
\vec{k}^2}+\frac{m_\sigma^2\overline{\sigma}^2}{2 \rho_B}+
\frac{g_\omega^2\rho_B}{2m_\omega^2} .
\label{toten}
\ene

We determine the coupling constants, $g_{\sigma}$ and $g_{\omega}$, so as 
to fit the binding energy ($-15.7$ MeV) per nucleon and the saturation 
density, $\rho_0$ = 0.15 fm$^{-3}$ ($k_F^0$ = 1.305 fm$^{-1}$), for 
symmetric nuclear matter.  
The coupling constants and some calculated properties 
of nuclear matter (for $m_{\sigma}=550$ MeV and 
$m_{\omega}=783$ MeV) at the saturation density are listed in 
Table~\ref{c.c.}. 
The last three columns show the relative changes (from their values at 
zero density) of the bag radius 
($\delta R_B^{\star}/R_B$), the lowest eigenvalue ($\delta x/x_0$) 
and the rms radius of the nucleon calculated by the quark wave function 
($\delta r_q^{\star}/r_q$) at saturation density. 
\begin{table}[htbp]
\begin{center}
\caption{Coupling constants and calculated properties for 
symmetric nuclear matter at normal nuclear density.  
The effective nucleon mass, $M_N^{\star}$, and the nuclear 
incompressibility, $K$, are quoted in MeV. 
The bottom row is for QHD.}
\label{c.c.}
\begin{tabular}[t]{c|ccccccccc}
\hline
$m_q$(MeV)&$R_B$(fm)&$g_{\sigma}^2/4\pi$&$g_{\omega}^2/4\pi$&
$M_N^{\star}$&$K$&$\delta R_B^{\star}/R_B$&$\delta x/x_0$&$\delta 
r_q^{\star}/r_q$ \\
\hline
   &  0.6 & 5.84 & 6.29 & 730 & 293 & $-0.02$ & $-0.13$ & 0.01 \\
 0 &  0.8 & 5.38 & 5.26 & 756 & 278 & $-0.02$ & $-0.16$ & 0.02 \\
   &  1.0 & 5.04 & 4.50 & 774 & 266 & $-0.02$ & $-0.19$ & 0.02 \\
\hline
   &  0.6 & 5.86 & 6.34 & 729 & 295 & $-0.02$ & $-0.13$ & 0.01 \\
 5 &  0.8 & 5.40 & 5.31 & 754 & 280 & $-0.02$ & $-0.16$ & 0.02 \\
   &  1.0 & 5.07 & 4.56 & 773 & 267 & $-0.02$ & $-0.19$ & 0.02 \\
\hline
   &  0.6 & 5.87 & 6.37 & 728 & 295 & $-0.02$ & $-0.13$ & 0.02 \\
10 &  0.8 & 5.42 & 5.36 & 753 & 281 & $-0.02$ & $-0.16$ & 0.02 \\
   &  1.0 & 5.09 & 4.62 & 772 & 269 & $-0.02$ & $-0.18$ & 0.03 \\
\hline
 QHD &    & 7.29 & 10.8 & 522 & 540 & & &  \\
\hline
\end{tabular}
\end{center}
\end{table}

The most notable fact is that the calculated incompressibility, $K$, 
is well within the experimental range: $K = 200 \sim 
300$ MeV.   Also our effective nucleon mass is much larger than in the 
case of QHD.  
Although the bag radius shrinks a little at finite density, the rms radius 
of the quark wavefunction actually {\em increases} by a few percent 
at saturation density. 

In Fig.\ref{fig:mass}, we show the  
effective nucleon mass in medium, as a function of density, 
for $m_q$ = 5 MeV. 
Its dependence on the quark mass is weak. 
The scalar density ratio $C({\overline \sigma})$ 
decreases linearly (to a very good approximation) with 
$g_{\sigma} {\overline \sigma}$\cite{finite}.  
It is very useful to have a simple parametrization for $C$ and 
the form:
\bge
C({\overline \sigma}) = 1 - a \times (g_{\sigma}{\overline \sigma}) , 
\label{paramC}
\ene
with $g_{\sigma} {\overline \sigma}$ in MeV (recall $g_{\sigma} = 
g_{\sigma}(\sigma=0)$) and 
$a$ = (6.6, 8.8, 11) $\times 10^{-4}$ for $R_B$ = (0.6, 0.8, 1.0) fm, 
respectively, is quite accurate up to $2 \sim 3\rho_0$. 
(These values for the slope parameter, $a$, are valid for 
all quark masses listed.)  

As a practical matter, it is easy to solve Eq.(\ref{deriv2}) for 
$g_\sigma(\sigma)$ in the 
case where $C(\sigma)$ is linear in $g_{\sigma} \overline{\sigma}$ as 
Eq.(\ref{paramC}).  In fact, we find 
\bge
M^{\star}_N = M_N - g_\sigma \left[ 1 - \frac{a}{2} (g_\sigma 
{\overline \sigma}) \right] {\overline \sigma} . 
\label{mstaR}
\ene
The effective $\sigma$-N coupling constant, $g_{\sigma}(\sigma)$, 
decreases at half the rate of $C(\sigma)$.  
Eq.(\ref{mstaR}) is quite accurate up to twice nuclear matter density.  

Having explicitly solved the nuclear matter problem by 
self-consistently solving for the quark wave functions in the bag in the 
mean scalar field, {\em one can solve for the properties of finite nuclei 
without explicit reference} to the internal structure of the nucleon. 
All one needs is Eqs.(\ref{paramC}) and (\ref{mstaR}) for $C(\sigma)$ and 
$M^{\star}_N$ as a function of $g_{\sigma} \overline{\sigma}$.

\clearpage

\subsection{Finite nuclei}
\label{fnt}
To describe a nucleus with different numbers of protons and neutrons 
($Z \neq N$), it is necessary to consider the contributions of the $\rho$
meson. Any realistic treatment of nuclear structure also requires that one
introduces the Coulomb force.
The interaction Lagrangian density for the $\rho$- and 
$\gamma$-quark couplings in the nucleon IRF is given by 
\bge
{\cal L}_I^{\rho+\gamma} = - g^q_{\rho} \bar{q}' \frac{\tau_{\alpha}^q}{2} 
\gamma_{\mu} q' \rho^{\mu}_{\alpha,IRF} - e \bar{q}' (\frac{1}{6} + 
\frac{\tau_3^q}{2}) \gamma_{\mu} q' A^{\mu}_{IRF} , 
\label{rhogamma}
\ene
where $q$ is the quark field (the prime means the IRF) and 
$\rho^\mu_{\alpha,IRF}$ and $A^{\mu}_{IRF}$ 
are respectively the $\rho$ meson field with isospin component $\alpha$ 
and the photon field.  
$\tau_{\alpha}^q/2$ is the isospin operator for the quarks.  $g_{\rho}^q$ 
and $e$ are the quark-$\rho$ coupling constant and the electric charge. 

We saw in Ref.\cite{finite} that this leads to both a central and a
spin-orbit potential for the nucleon in the nuclear rest frame. For the
isoscalar, $\omega$ meson these potentials are well represented by a
vector $\omega$-nucleon coupling because the isoscalar magnetic moment
of the nucleon is near unity. However, for the $\rho$, the relativistic
formulation {\em at the nucleon level} requires a strong tensor coupling
($\sim \sigma^{\mu \nu} q_\nu$) if it is to reproduce the interaction
given by the quark model\cite{finite}.

In Hartree approximation a tensor coupling gives only a spin-orbit force
for a nucleon bound in a static, spherical nucleus. However, in
Hartree-Fock it can also give rise to a central force which would be
necessary to reproduce the bulk symmetry energy. For simplicity, in the
present work we follow the usual procedure adopted in the Hartree
treatment of QHD of using only a vector coupling for the $\rho$, with
$g_\rho$ adjusted to give the bulk symmetry energy in MFA -- where
$g_\rho = g_\rho^q$. (The accuracy
of this approximation will be examined in detail in future work.) 

In this case, in mean field approximation, only $\alpha =3$ contributes
in Eq.(\ref{rhogamma}). 
If we denote by $b(\vec{r})$ and $A({\vec r})$ the mean values of the 
time component of $\rho$ field and the Coulomb field in the nuclear 
rest frame, 
we can transpose our results for the $\omega$ field by trivial isospin 
factors:
\bge
3g^q_\omega \omega({\vec r}) \to g_\rho\frac{\tau^N_3}{2} b({\vec r}) 
\ \ \mbox{or} \ \ \frac{e}{2}(1 + \tau^N_3) A({\vec r}) ,
\label{subst}
\ene
where $\tau^N_3/2$ is the third component of the nucleon 
isospin operator. 

In summary, our effective Lagrangian density in mean field 
approximation takes the form:
\bg
{\cal L}_{QMC}&=& \overline{\psi} [i \gamma \cdot \partial 
- M_N  + g_\sigma (\sigma({\vec r})) \sigma({\vec r}) 
- g_\omega \omega({\vec r}) \gamma_0 \nn \\
&-& g_\rho \frac{\tau^N_3}{2} b({\vec r}) \gamma_0 
- \frac{e}{2} (1+\tau^N_3) A({\vec r}) \gamma_0 ] \psi \nn \\
&-& \frac{1}{2}[ (\nabla \sigma({\vec r}))^2 + 
m_{\sigma}^2 \sigma({\vec r})^2 ] 
+ \frac{1}{2}[ (\nabla \omega({\vec r}))^2 + m_{\omega}^2 
\omega({\vec r})^2 ] \nn \\
&+& \frac{1}{2}[ (\nabla b({\vec r}))^2 + m_{\rho}^2 b({\vec r})^2 ] 
+ \frac{1}{2} (\nabla A({\vec r}))^2. 
\label{qmclag}
\en
The variation of the Lagrangian results in the 
following equations for static, spherically symmetric nuclei (see also 
Eqs.(\ref{omg}) and (\ref{sig2})): 
\bg
\frac{d^2}{dr^2} \sigma(r) + \frac{2}{r} \frac{d}{dr} \sigma(r) 
    - m_\sigma^2 \sigma(r) &=& - g_\sigma C(\sigma(r)) \rho_s(r) \nn \\
    &\equiv& - g_\sigma C(\sigma(r)) \sum_\alpha^{occ} d_\alpha(r) 
    (|G_\alpha(r)|^2 - |F_\alpha(r)|^2), \label{sig3} \\
\frac{d^2}{dr^2} \omega(r) + \frac{2}{r} \frac{d}{dr} \omega(r) 
    - m_\omega^2 \omega(r) &=& - g_\omega \rho_B(r) \nn \\
    &\equiv& - g_\omega \sum_\alpha^{occ} d_\alpha(r)
    (|G_\alpha(r)|^2 + |F_\alpha(r)|^2), \label{omg3} \\
\frac{d^2}{dr^2} b(r) + \frac{2}{r} \frac{d}{dr} b(r) 
    - m_\rho^2 b(r) &=& - \frac{g_\rho}{2} \rho_3(r) \nn \\
    &\equiv& - \frac{g_\rho}{2} \sum_\alpha^{occ} 
    d_\alpha(r) (-)^{t_\alpha -1/2} 
    (|G_\alpha(r)|^2 + |F_\alpha(r)|^2), \label{rho3} \\
\frac{d^2}{dr^2} A(r) + \frac{2}{r} \frac{d}{dr} A(r) 
    &=& - e \rho_p(r) \nn \\
    &\equiv& - e \sum_\alpha^{occ} d_\alpha(r) 
    (t_\alpha + \frac{1}{2}) 
    (|G_\alpha(r)|^2 + |F_\alpha(r)|^2), \label{phtn3} 
\en
where $d_\alpha(r)= (2j_\alpha+1)/4\pi r^2$ and 
\bg
\frac{d}{dr} G_\alpha(r) + \frac{\kappa}{r} G_\alpha(r) - 
\left[ \epsilon_\alpha - g_\omega \omega(r) - t_\alpha g_\rho b(r)
\right. 
&-& \left. (t_\alpha + \frac{1}{2}) e A(r) + M_N \right. \nn \\
&-& \left. g_\sigma(\sigma(r)) \sigma(r) \right] F_\alpha(r) = 0 , 
\label{qwave1} \\
\frac{d}{dr} F_\alpha(r) - \frac{\kappa}{r} F_\alpha(r) + 
\left[ \epsilon_\alpha - g_\omega \omega(r) - t_\alpha g_\rho b(r)
\right.
&-& \left. (t_\alpha + \frac{1}{2}) e A(r) - M_N \right. \nn \\
&+& \left. g_\sigma (\sigma(r)) \sigma (r) \right] G_\alpha(r) = 0 . 
\label{qwave2} 
\en
Here $iG_\alpha(r)/r$ and $-F_\alpha(r)/r$ are respectively the 
radial part of the upper and lower 
components of the solution to the Dirac equation for the nucleon 
($\alpha$ labelling the quantum numbers and $\epsilon_\alpha$ being 
the energy) under the normalization condition: 
\bge
\int dr (|G_\alpha(r)|^2 + |F_\alpha(r)|^2) =1 . \label{norm}
\ene
As usual, $\kappa$ specifies the angular quantum numbers and $t_\alpha$ 
the eigenvalue of the isospin operator $\tau^N_3/2$.  $C(\sigma)$ and 
$g_\sigma (\sigma)$ are given by Eqs.(\ref{sss}) and 
(\ref{coup}), respectively -- or practically, 
by Eqs.(\ref{paramC}) and (\ref{mstaR}), i.e., 
\bge
g_\sigma (\sigma({\vec r})) =  g_\sigma \left[ 1 - \frac{a}{2} 
            g_\sigma \sigma({\vec r}) \right] . \label{gsgm}
\ene
The total energy of the system is then given by 
\bg
E_{tot} &=& \sum_\alpha^{occ} (2j_\alpha + 1) \epsilon_\alpha 
 - \frac{1}{2} \int d{\vec r} \ [ -g_\sigma C(\sigma (r)) \sigma(r) 
\rho_s(r) \nn \\
 &+& g_\omega \omega(r) \rho_B(r) + \frac{1}{2} g_\rho b(r) \rho_3(r) 
 + eA(r) \rho_p(r) ].  \label{ftoten}
\en

\section{Numerical results for finite nuclei}
\label{results}
Equations (\ref{sig3}) to (\ref{norm}) give a set of coupled non-linear 
differential equations, which may be solved by a standard iteration 
procedure\cite{horowitz}.  In this paper the numerical calculation was 
carried out by modifying the technique described by Horowitz 
{\it et al.}\cite{horowitz,hor2}.  Changing the initial condition for 
the meson fields, the $\sigma$-N coupling constant and the scalar density 
in their fortran program\cite{hor2}, the calculation can be performed 
very easily\footnote{If the numerical convergence is slow 
it may be improved by mixing appropriately the meson potentials 
given by the $i$-th iteration and those by the ($i-1$)-th iteration -- as is 
usually done in non-relativistic calculations.}.  
The calculation is achieved in at most 
20 iterations when it is performed with a maximum radius of 12 (15) fm on 
a mesh of 0.04 fm for medium mass (Pb) nuclei.  

\subsection{Determination of parameters}
\label{param}
There are seven parameters to be determined: $g_\sigma$, $g_\omega$, 
$g_\rho$, $e$, $m_\sigma$, $m_\omega$ and $m_\rho$.  We take the 
experimental values: $m_\omega$ = 783 MeV, $m_\rho$ = 770 MeV and 
$e^2/4\pi$ = 1/137.036.  
The coupling constants $g_\sigma$ and 
$g_\omega$ are fixed to describe the nuclear matter properties 
with $m_\sigma$ = 550 MeV in Sec.\ref{matter} (see Table~\ref{c.c.}).  

The $\sigma$-meson mass however determines the range of the attractive 
interaction and changes in $m_\sigma$ affect the nuclear-surface slope and 
its thickness.  Therefore, as in the paper of Horowitz and 
Serot\cite{horowitz}, we adjust $m_\sigma$ to fit the rms charge radius of 
$^{40}$Ca, $r_{ch}$($^{40}$Ca) = 3.48 fm, the experimental 
value\cite{sinha}.  
We should notice here that {\em variations of $m_\sigma$ at fixed} 
($g_\sigma / m_\sigma$) 
{\em have no effect on the infinite nuclear matter properties}.  Therefore, 
keeping the ratio ($g_\sigma / m_\sigma$) constant we vary $m_\sigma$ to fit 
the rms charge radius of $^{40}$Ca. We 
expect that $m_\sigma$ ranges around 400 $\sim$ 550 MeV\cite{blun,mach}.  
The last parameter, $g_\rho$, is adjusted to yield the bulk symmetry 
energy per 
baryon of 35 MeV\cite{horowitz}.  We summarize the parameters in 
Table~\ref{c.c.2}.
\begin{table}[htbp]
\begin{center}
\caption{Model parameters for finite nuclei.}
\label{c.c.2}
\begin{tabular}[t]{cccccc}
\hline
$m_q$(MeV)&$R_B$(fm)&$g_{\sigma}^2/4\pi$&$g_{\omega}^2/4\pi$&
$g_{\rho}^2/4\pi$&$m_\sigma$(MeV) \\
\hline
   &  0.6 & 3.55 & 6.29 & 6.79 & 429 \\
 0 &  0.8 & 2.94 & 5.26 & 6.93 & 407 \\
   &  1.0 & 2.51 & 4.50 & 7.03 & 388 \\
\hline
   &  0.6 & 3.68 & 6.34 & 6.78 & 436 \\
 5 &  0.8 & 3.12 & 5.31 & 6.93 & 418 \\
   &  1.0 & 2.69 & 4.56 & 7.02 & 401 \\
\hline
   &  0.6 & 3.81 & 6.37 & 6.78 & 443 \\
10 &  0.8 & 3.28 & 5.36 & 6.92 & 428 \\
   &  1.0 & 2.91 & 4.62 & 7.02 & 416 \\
\hline
\end{tabular}
\end{center}
\end{table}
\begin{figure}[htb]
\centering{\
\epsfig{file=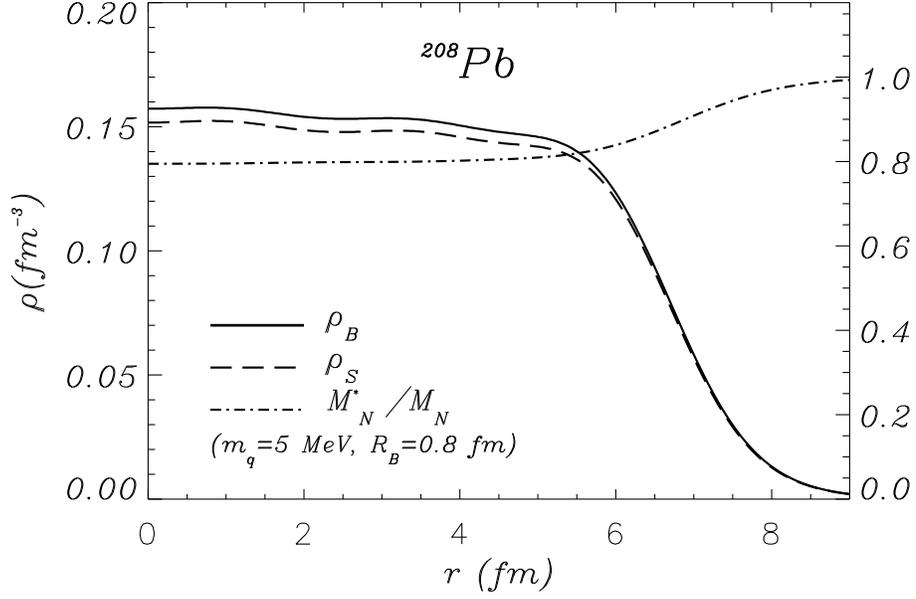,height=9cm}
\caption{Model predictions for the effective nucleon mass, the baryon and 
scalar densities in $^{208}$Pb (for $m_q$ = 5 MeV and $R_B$ = 0.8 fm).  
The scale on the right vertical axis is for $M_N^{\star}/M_N$. }
\label{pbd58}}
\end{figure}
In Fig.\ref{pbd58}, we show the baryon and scalar densities as well as 
the effective nucleon mass in lead.  We expect that the baryon density 
in the interior of lead would be close to the saturation density of infinite 
nuclear matter.  As seen in the figure, the calculated baryon density  
at the center is quite close to 0.15 fm$^{-1}$, which supports our choice of 
the parameters.  
\clearpage

\subsection{Comparison with experimental data}
\label{comp}
\begin{figure}[htb]
\centering{\
\epsfig{file=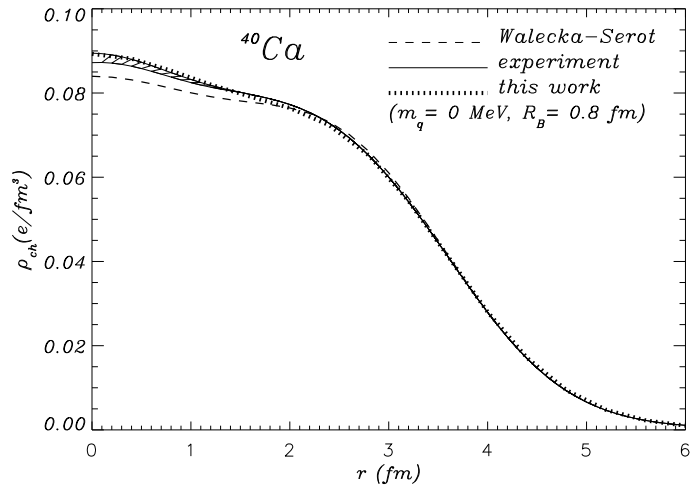,height=7cm}
\caption{Charge density distribution for $^{40}$Ca (for $m_q$ = 0 MeV and 
$R_B$ = 0.8 fm) compared with the experimental data and that of QHD. }
\label{ca08}}
\end{figure}

\begin{figure}[htb]
\centering{\
\epsfig{file=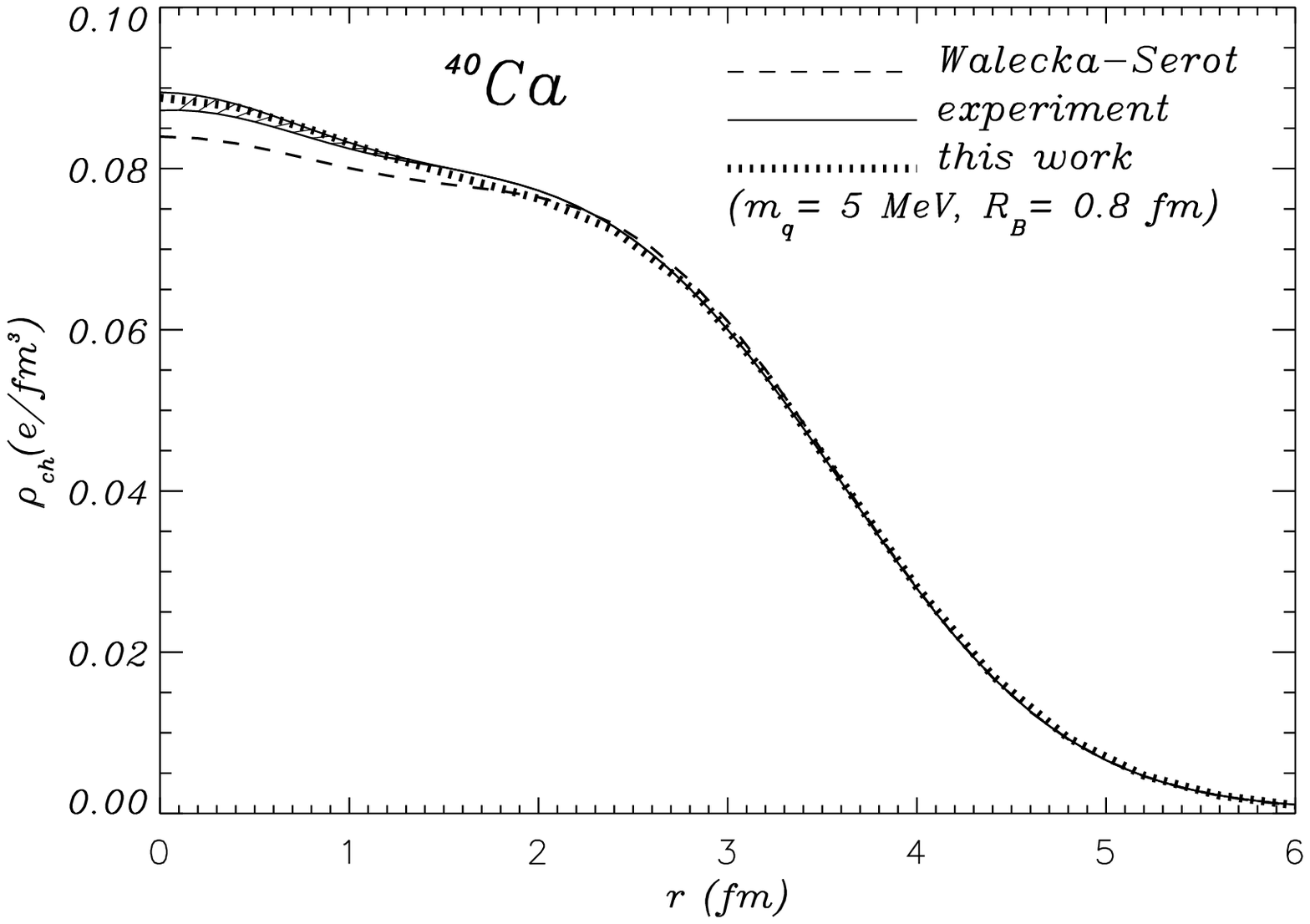,height=7cm}
\caption{Same as Fig.5 (for $m_q$ = 5 MeV and $R_B$ = 0.8 fm). }
\label{ca58}}
\end{figure}

\begin{figure}[htb]
\centering{\
\epsfig{file=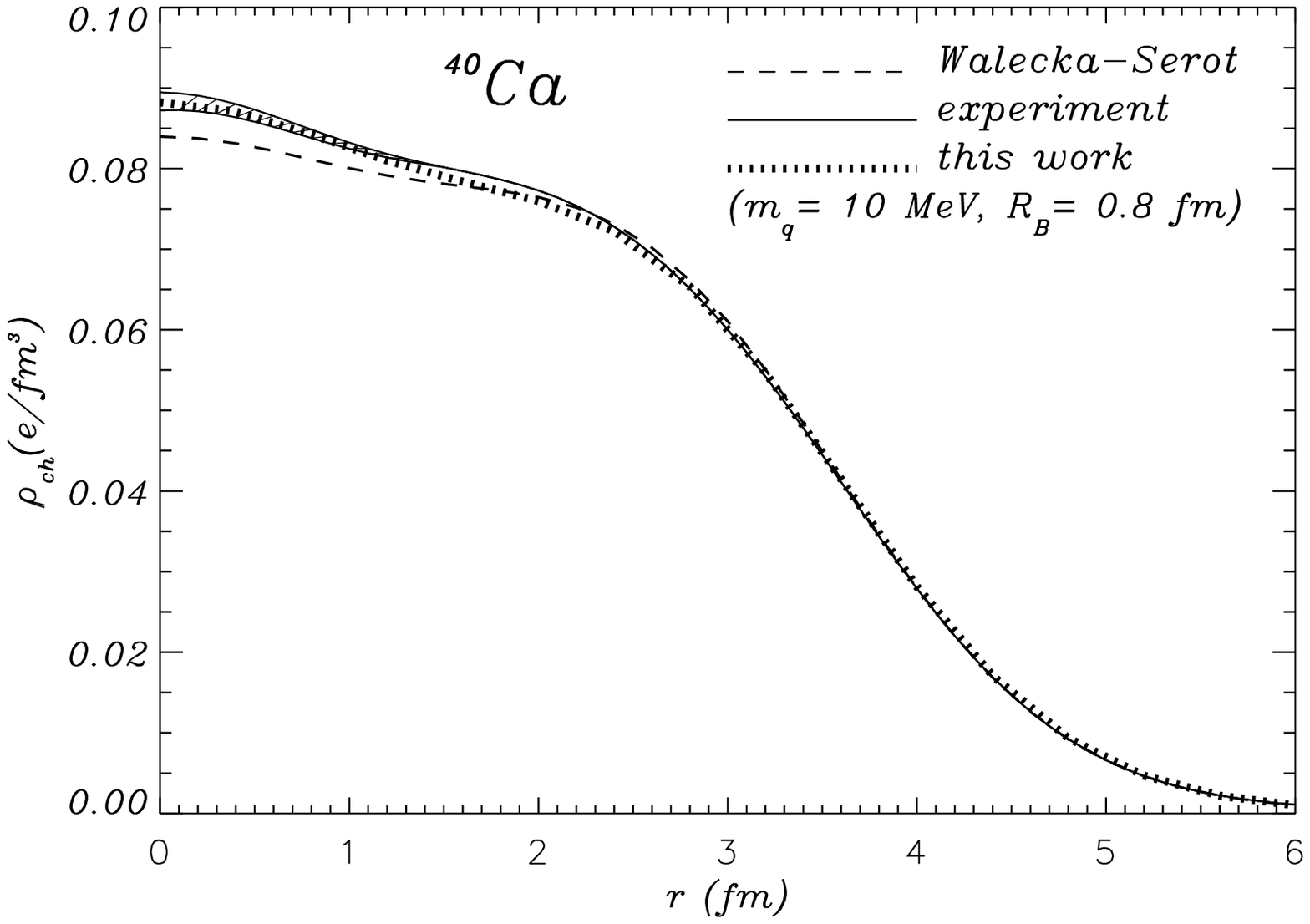,height=7cm}
\caption{Same as Fig.5 (for $m_q$ = 10 MeV and $R_B$ = 0.8 fm). }
\label{ca18}}
\end{figure}

\begin{figure}[htb]
\centering{\
\epsfig{file=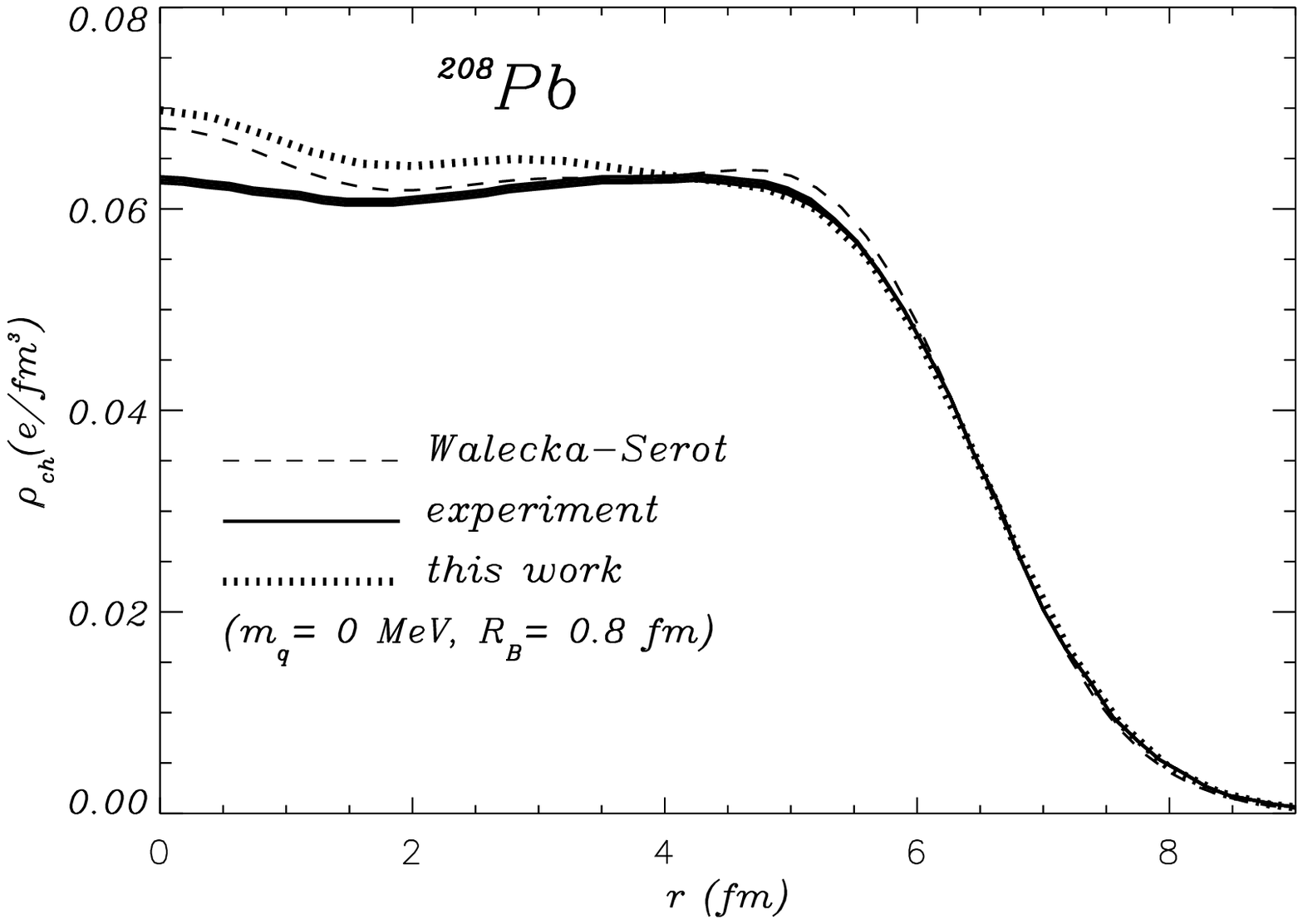,height=7cm}
\caption{Charge density distribution for $^{208}$Pb (for $m_q$ = 0 MeV and 
$R_B$ = 0.8 fm) compared with the experimental data and that of QHD. }
\label{pb08}}
\end{figure}

\begin{figure}[htb]
\centering{\
\epsfig{file=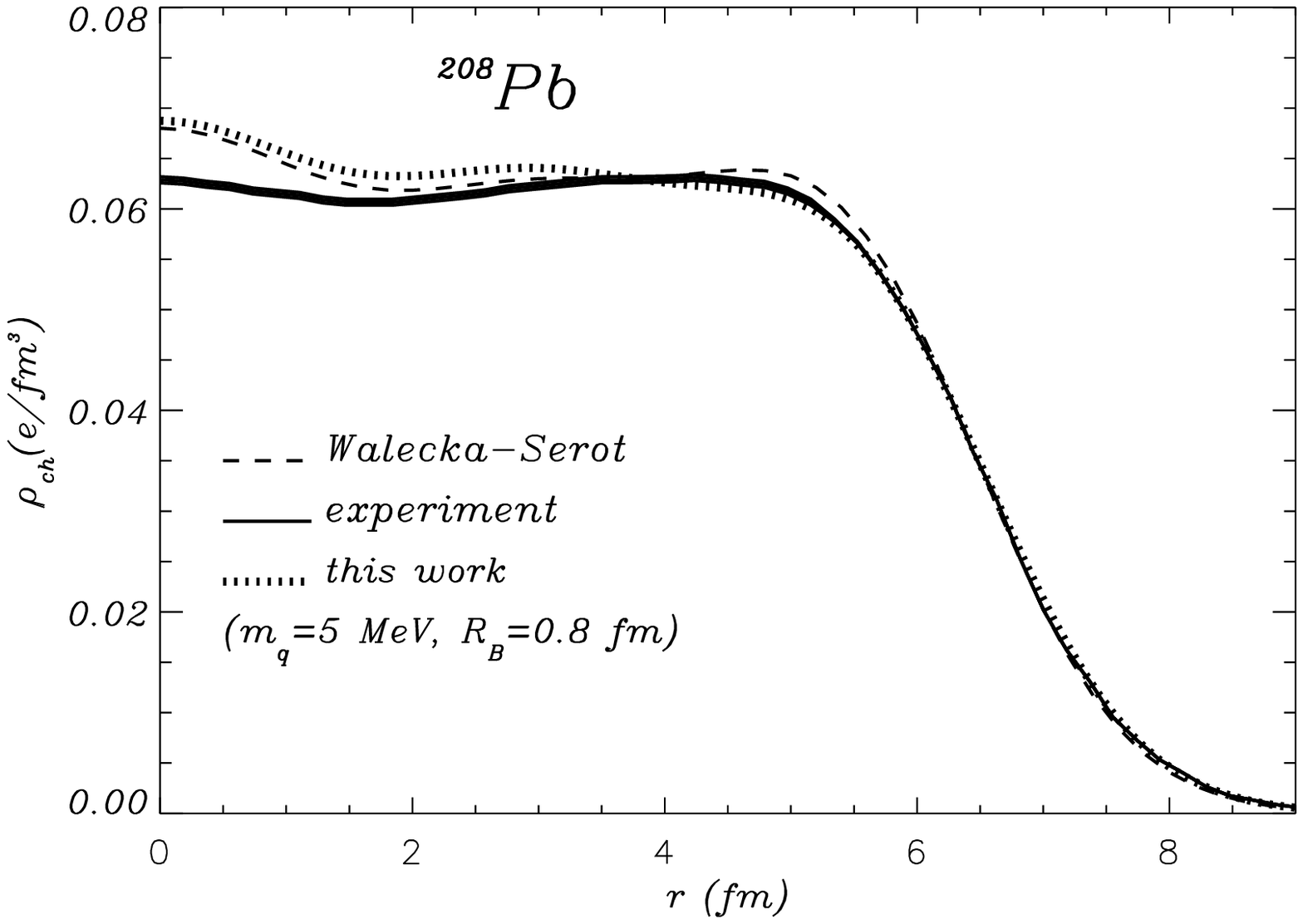,height=7cm}
\caption{Same as Fig.8 (for $m_q$ = 5 MeV and $R_B$ = 0.8 fm). }
\label{pb58}}
\end{figure}

\begin{figure}[htb]
\centering{\
\epsfig{file=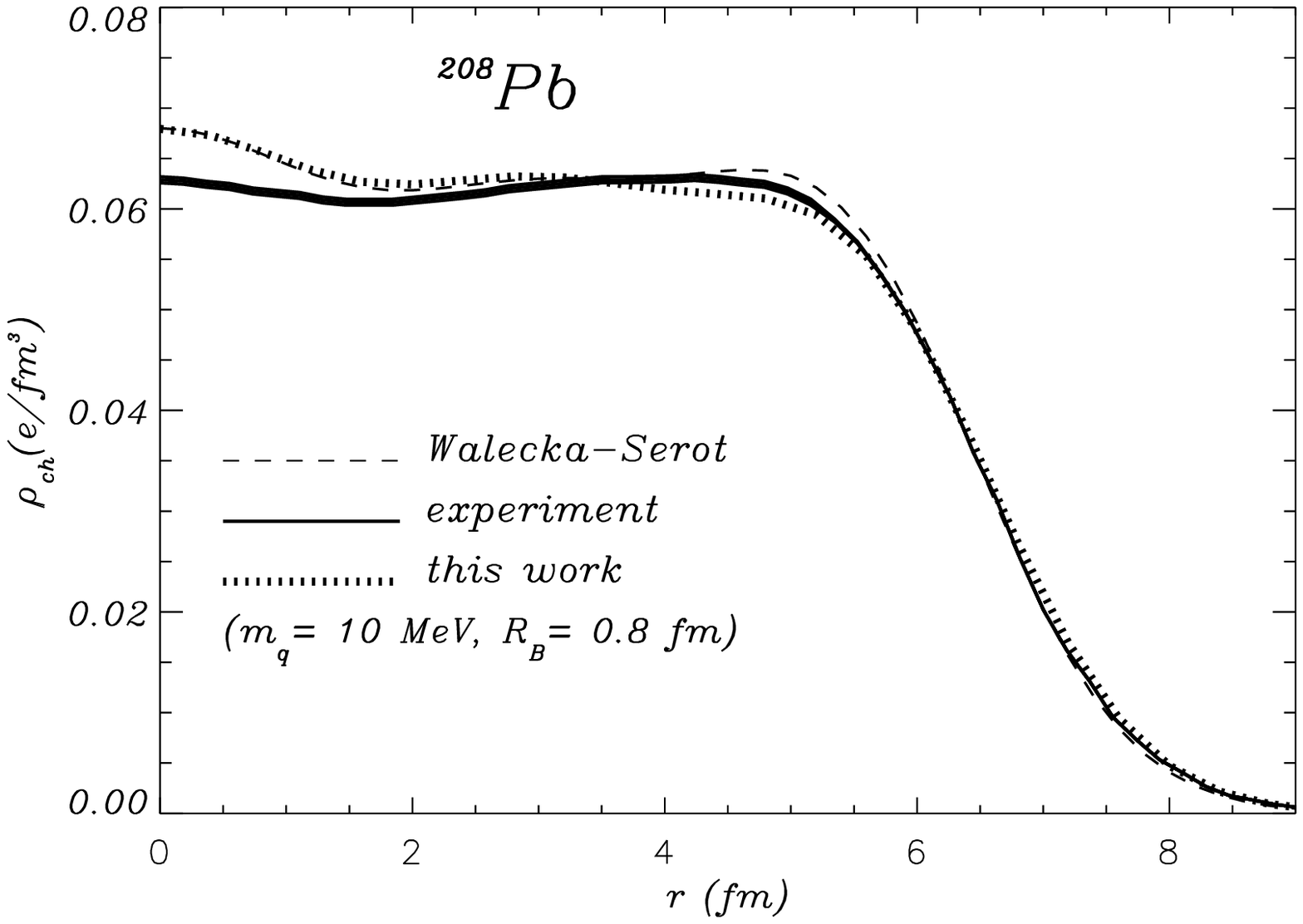,height=7cm}
\caption{Same as Fig.8 (for $m_q$ = 10 MeV and $R_B$ = 1.0 fm). }
\label{pb18}}
\end{figure}

\begin{figure}[htb]
\centering{\
\epsfig{file=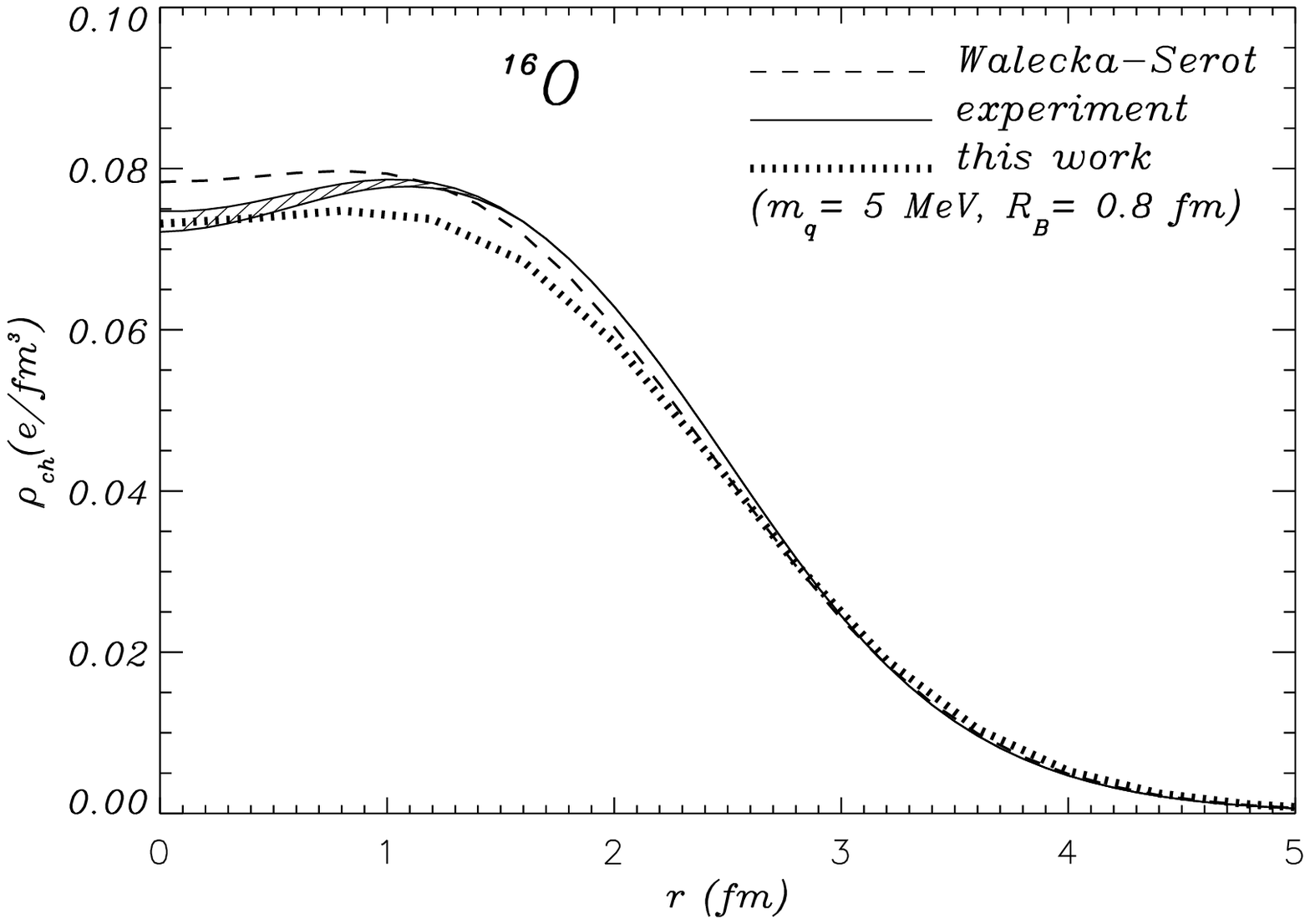,height=7cm}
\caption{Charge density distribution for $^{16}$O (for $m_q$ = 5 MeV and 
$R_B$ = 0.8 fm) compared with the experimental data and that of QHD. }
\label{o58}}
\end{figure}

\begin{figure}[htb]
\centering{\
\epsfig{file=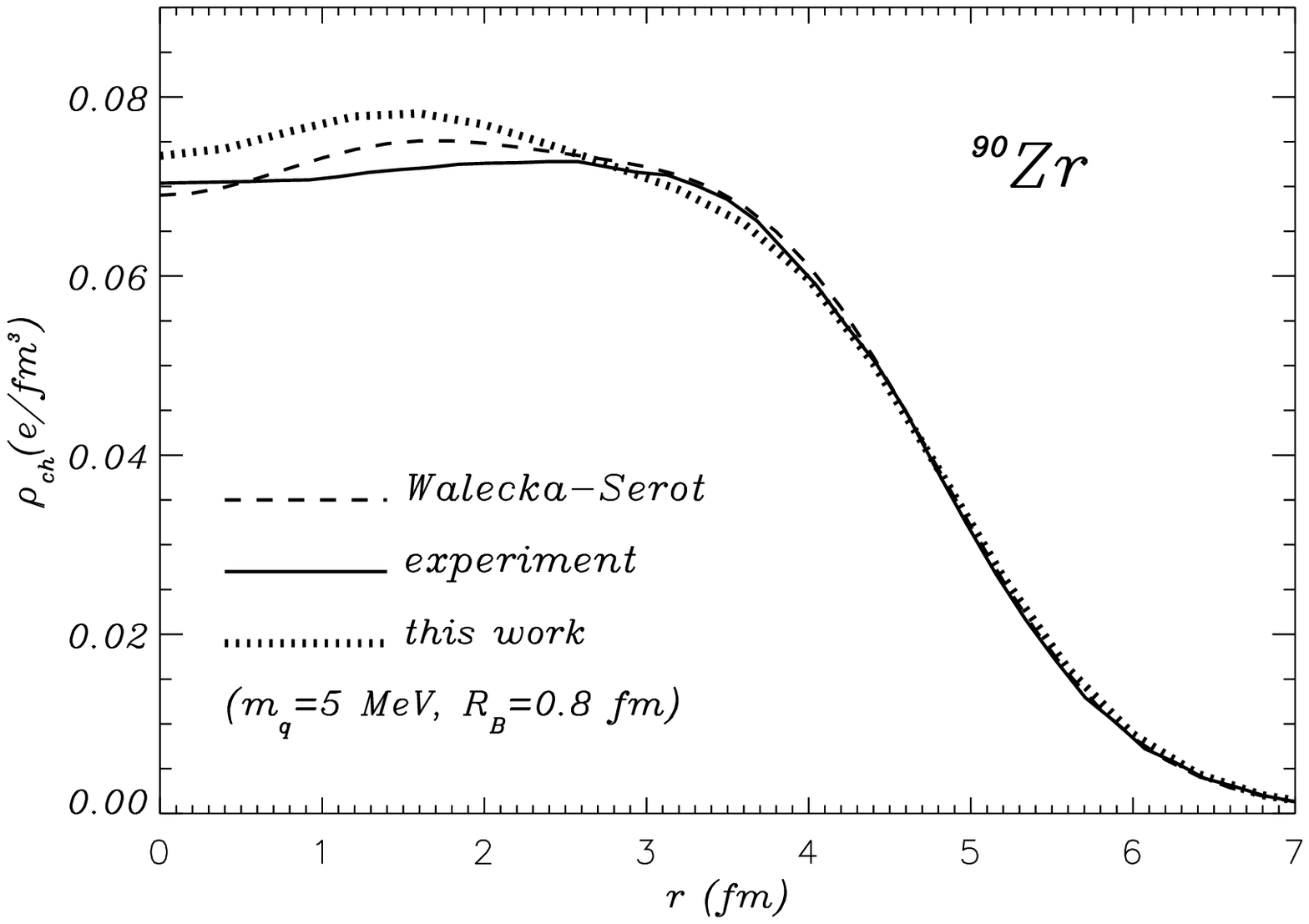,height=7cm}
\caption{Charge density distribution for $^{90}$Zr (for $m_q$ = 5 MeV and 
$R_B$ = 0.8 fm) compared with the experimental data and that of QHD. }
\label{zr58}}
\end{figure}

\begin{figure}[htb]
\centering{\
\epsfig{file=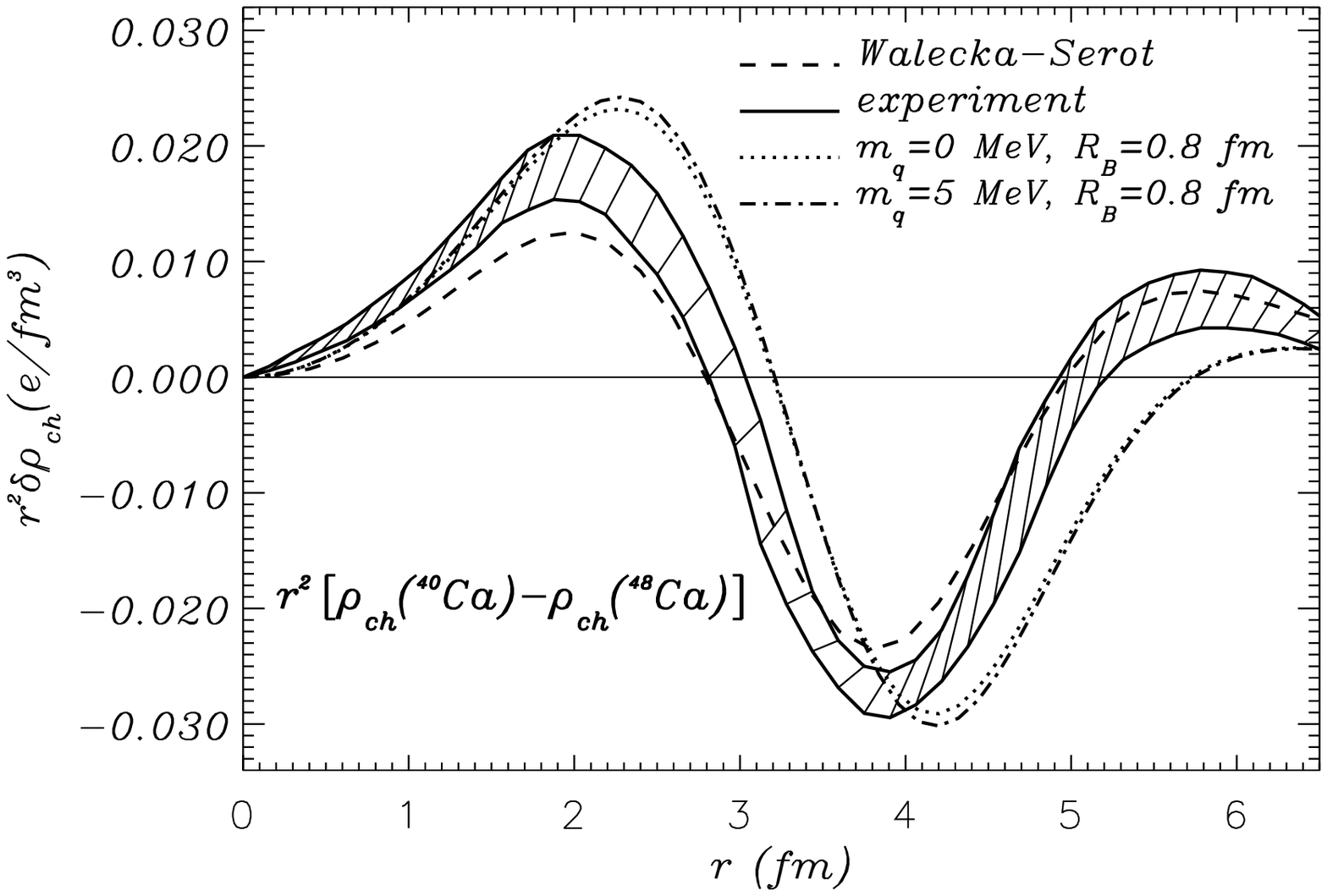,height=7cm}
\caption{Isotope shift between $\rho_{ch}$($^{40}$Ca) and 
$\rho_{ch}$($^{48}$Ca) compared with the experimental data and that of QHD 
(for $m_q$ = 0 and 5 MeV with $R_B$ = 0.8 fm).  }
\label{ch48a05}}
\end{figure}

\begin{figure}[htb]
\centering{\
\epsfig{file=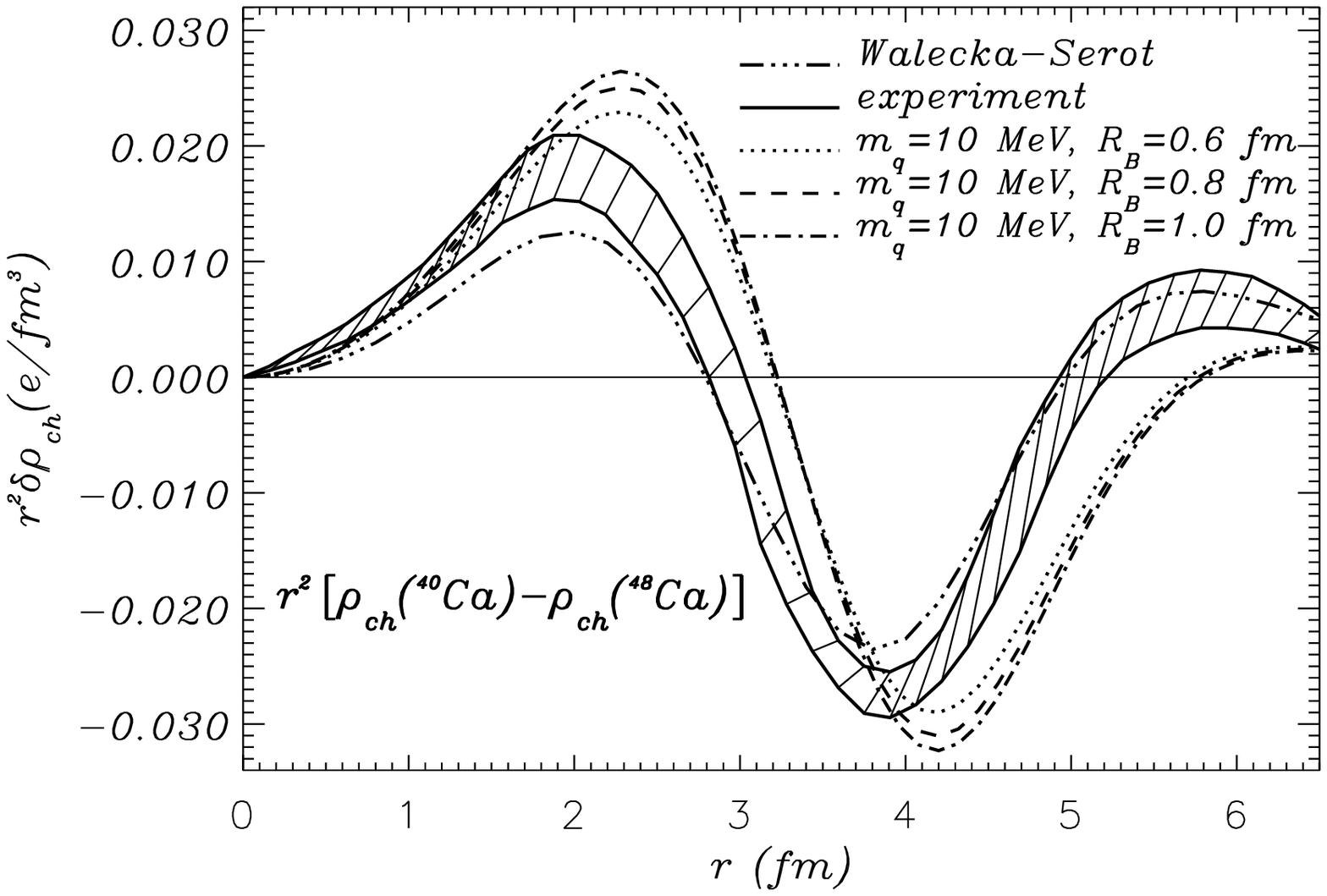,height=7cm}
\caption{Same as Fig.13 (for $m_q$ = 10 with $R_B$ = 0.6, 0.8 and 1.0 fm).}
\label{ch48a1}}
\end{figure}

First we show calculated charge density distributions, $\rho_{ch}$, 
in comparison with those of QHD\cite{qhd,horowitz} and the experimental 
data in Figs.\ref{ca08}$-$\ref{ch48a1}.  
Having solved Eqs.(\ref{sig3}) $\sim$ (\ref{norm}), we obtain the 
{\em point-}proton and neutron densities in a nucleus.  
In addition to those densities, we should include 
the effects of c.m. corrections, nucleon form factors, meson-exchange 
currents, etc.  However, one knows that the dominant correction comes 
from the proton-form factor\cite{horowitz}.   We calculate the charge
density as a 
convolution of the point-proton density, $\rho_p({\vec r})$, with 
the proton charge 
distribution, $\rho^p_{ch}({\vec r})$: 
\bge
\rho_{ch}({\vec r}) = \int d{\vec r}\, ' \ \rho^p_{ch}({\vec r} - 
{\vec r}\, ') \rho_p({\vec r}\, ') , \label{charge}
\ene
where we have used a gaussian form for $\rho^p_{ch}$ 
\bge
\rho^p_{ch}({\vec r}) = (\beta/\pi)^{3/2} \exp (-\beta r^2) . 
\label{gauss}
\ene
The parameter $\beta$, which determines the proton size, is chosen so 
as to reproduce the experimental rms charge radius of the proton, 0.82 fm 
(i.e., $\beta$ = 2.231 fm$^{-2}$).  
As we noted before, in the present model the rms radius of the nucleon 
in nuclear matter increases a little.  However, this 
amount is quite small and it can be ignored in the numerical calculations
 of nuclear parameters.  

The charge density distributions for $^{40}$Ca (for $m_q$ = 0, 5, 10 MeV 
and $R_B$ = 0.8 fm) are presented in Figs. \ref{ca08}$-$\ref{ca18}.  
The experimental data is taken from Ref.\cite{cadata}.  
Once the rms charge radius of $^{40}$Ca is fitted by adjusting $m_\sigma$ 
the QMC model can reproduce $\rho_{ch}$($^{40}$Ca) quite well.  
As seen in the figures, the calculated charge densities lie almost within 
the experimental (hatched) area.  We note that the dependence of 
$\rho_{ch}$($^{40}$Ca) on the bag radius is quite weak.  

It is interesting to see the quantum oscillations of the interior 
density in lead.  The dependence of $\rho_{ch}$($^{208}$Pb) on the 
quark mass is illustrated in Figs.\ref{pb08}$-$\ref{pb18}.  
We note that the dependence of $\rho_{ch}$($^{208}$Pb)
on the free nucleon size is also quite 
weak.  As seen in the figures, our model gives charge densities 
very close to those of QHD and still somewhat larger in the central 
region than those observed experimentally\cite{frois}.  

In Figs.\ref{o58} and \ref{zr58}, we show respectively the charge density 
distributions for $^{16}$O and $^{90}$Zr.  For zirconium the calculated 
$\rho_{ch}$ lies in between that of the non-relativistic density-dependent 
Hartree-Fock calculations\cite{nonrel} and that of QHD.  
The experimental data for oxygen and zirconium are taken 
from Refs.\cite{odata} and \cite{zdata}.  (For both cases the dependence of 
$\rho_{ch}$ on $m_q$ and $R_B$ is weak.)  

To see the isotope shift in charge density we have plotted $r^2$ times the 
difference between $\rho_{ch}$($^{40}$Ca) and $\rho_{ch}$($^{48}$Ca) in 
Figs.\ref{ch48a05} and \ref{ch48a1}.  Its dependence on the bag radius is 
again weak for a small quark mass while it becomes a little stronger 
for $m_q$ = 10 MeV, as shown in Fig.\ref{ch48a1}.  
The experimental data is taken from Ref.\cite{ch48data}. (Note that in
this case we also checked that including the charge distribution of the
neutron had a small effect.)

\begin{figure}[htb]
\centering{\
\epsfig{file=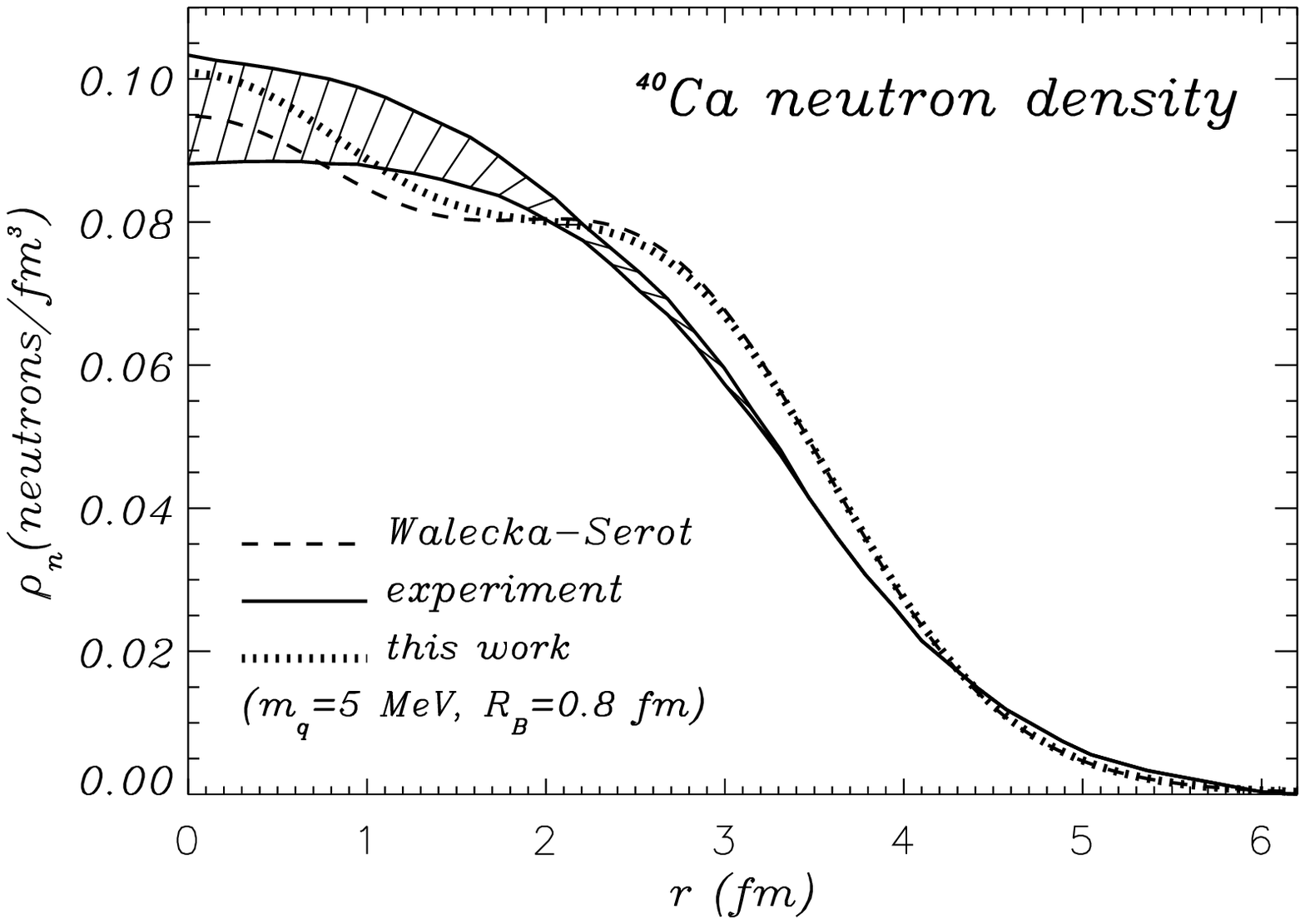,height=7cm}
\caption{Point-neutron density distribution in $^{40}$Ca 
(for $m_q$ = 5 MeV and 
$R_B$ = 0.8 fm) compared with that of QHD and the empirical fit. }
\label{can58}}
\end{figure}

\begin{figure}[htb]
\centering{\
\epsfig{file=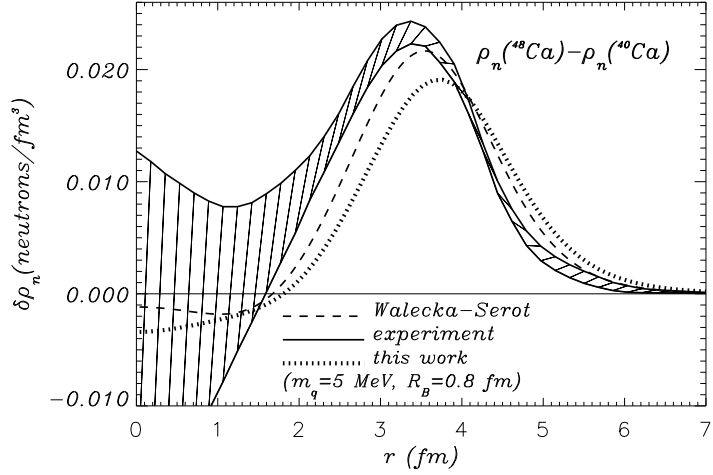,height=7cm}
\caption{Difference between $\rho_{n}$($^{48}$Ca) and $\rho_{n}$($^{40}$Ca) 
compared with that of QHD and the empirical fit 
(for $m_q$ = 5 and $R_B$ = 0.8 fm). }
\label{nd58}}
\end{figure}

\begin{figure}[htb]
\centering{\
\epsfig{file=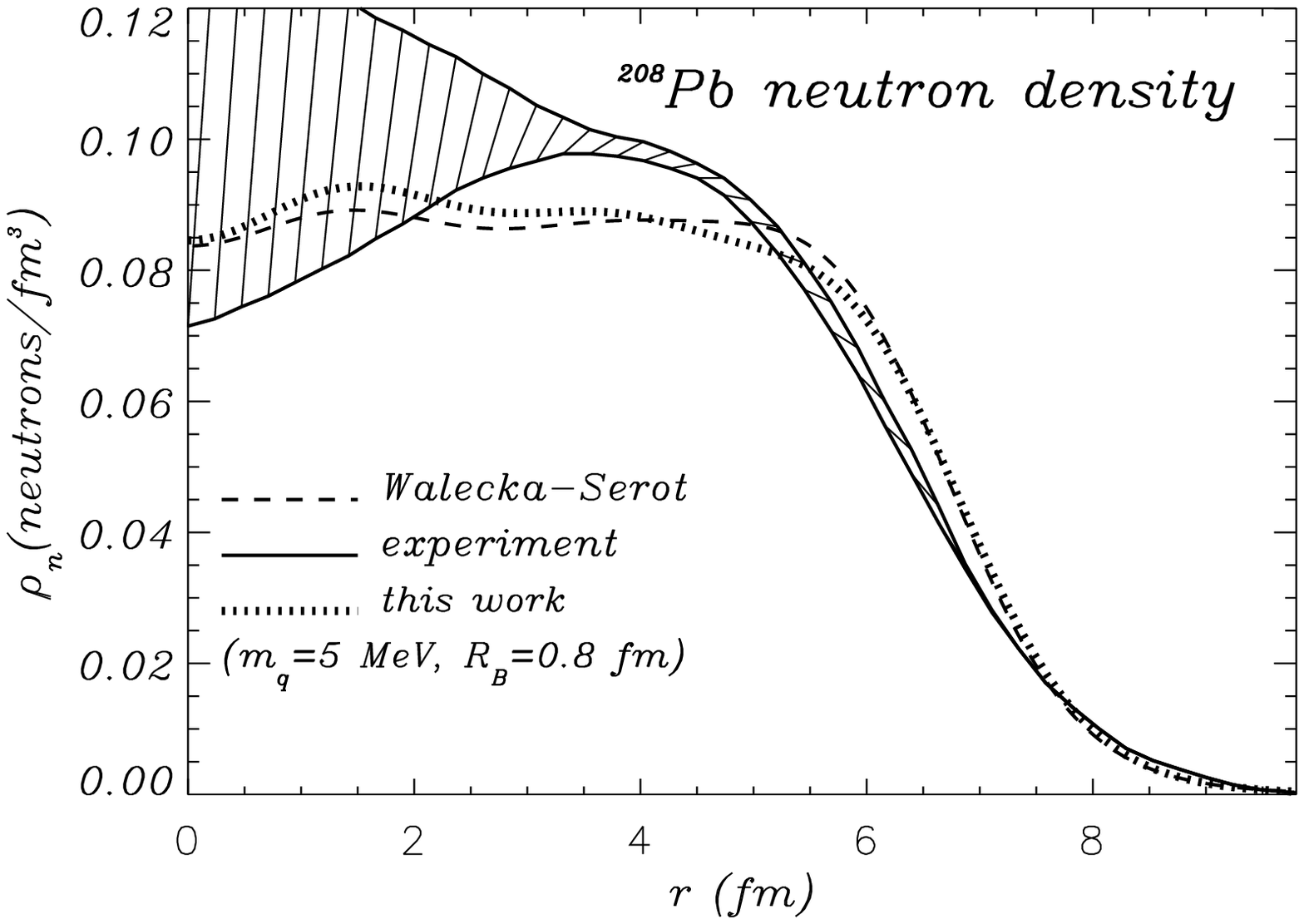,height=7cm}
\caption{Point-neutron density distribution in $^{208}$Pb 
(for $m_q$ = 5 MeV and $R_B$ = 0.8 fm) 
compared with that of QHD and the empirical fit. }
\label{pbn58}}
\end{figure}

In Figs.\ref{can58}$-$\ref{pbn58}, we present the point-neutron density 
distributions, $\rho_n$, in calcium and lead.  
For $^{40}$Ca, since the dependence of $\rho_n$ on $m_q$ and $R_B$ is again 
fairly weak, only the result for $m_q$ = 5 MeV and $R_B$ = 0.8 fm is 
shown, together with the empirical fit\cite{ncadata} to proton 
scattering data.  
We again find reasonable agreement with the data.  
For the isotope shift of $\rho_n$($^{48}$Ca)$-\rho_n$($^{40}$Ca), 
the calculated difference is closer to those of non-relativistic results 
than to those of QHD.  
The neutron density distribution in lead is shown in Fig.\ref{pbn58}.  
Its behavior is again similar to that of QHD.  

Table~\ref{sum} gives a summary of the calculated binding energy per nucleon 
($E/A$), rms charge radii and the difference between nuclear rms radii for 
neutrons and protons ($r_n - r_p$), for several closed-shell nuclei.  
\begin{table}[htbp]
\begin{center}
\caption{Binding energy per nucleon $E/A$ (in MeV), rms charge radius 
$r_{ch}$ (in fm) and difference between $r_n$ and $r_p$ (in fm).  
$m_q$ = 5 MeV and $R_B$ = 0.8 fm. ($^*$ fit) }
\label{sum}
\begin{tabular}[t]{c|ccc|ccc|ccc}
\hline
 & & $-E/A$ & & & $r_{ch}$ & & & $r_n-r_p$ & \\
\hline
Model & QMC & QHD & Exp. & QMC & QHD & Exp. & QMC & QHD & Exp. \\
\hline
$^{16}$O &5.84&4.89&7.98&2.79&2.75&2.73&$-0.03$&$-0.03$&0.0 \\
$^{40}$Ca&7.36&6.31&8.45&3.48$^*$&3.48$^*$&3.48&$-0.05$&$-0.06$&0.05$\pm$0.05\\
$^{48}$Ca&7.26&6.72&8.57&3.52&3.47&3.47&0.23&0.21&0.2$\pm$0.05 \\
$^{90}$Zr&7.79&7.02&8.66&4.27&4.26&4.27&0.11&0.10&0.05$\pm$0.1 \\
$^{208}$Pb&7.25&6.57&7.86&5.49&5.46&5.50&0.26&0.27&0.16$\pm$0.05 \\
\hline
\end{tabular}
\end{center}
\end{table}
Since the calculated properties do not depend strongly on $m_q$ and $R_B$, 
we only list the values for the QMC model with $m_q$ = 5 MeV 
and $R_B$ = 0.8 fm.  References for the experimental values can be found in 
Ref.\cite{horowitz}.  While there are still some discrepancies between 
the results and data, the present model provides quite reasonable results.  
In particular, it gives much larger binding energies per nucleon 
than those of 
QHD while still reproducing the rms charge radii for medium
and heavy nuclei quite well.  

\begin{figure}[htb]
\centering{\
\epsfig{file=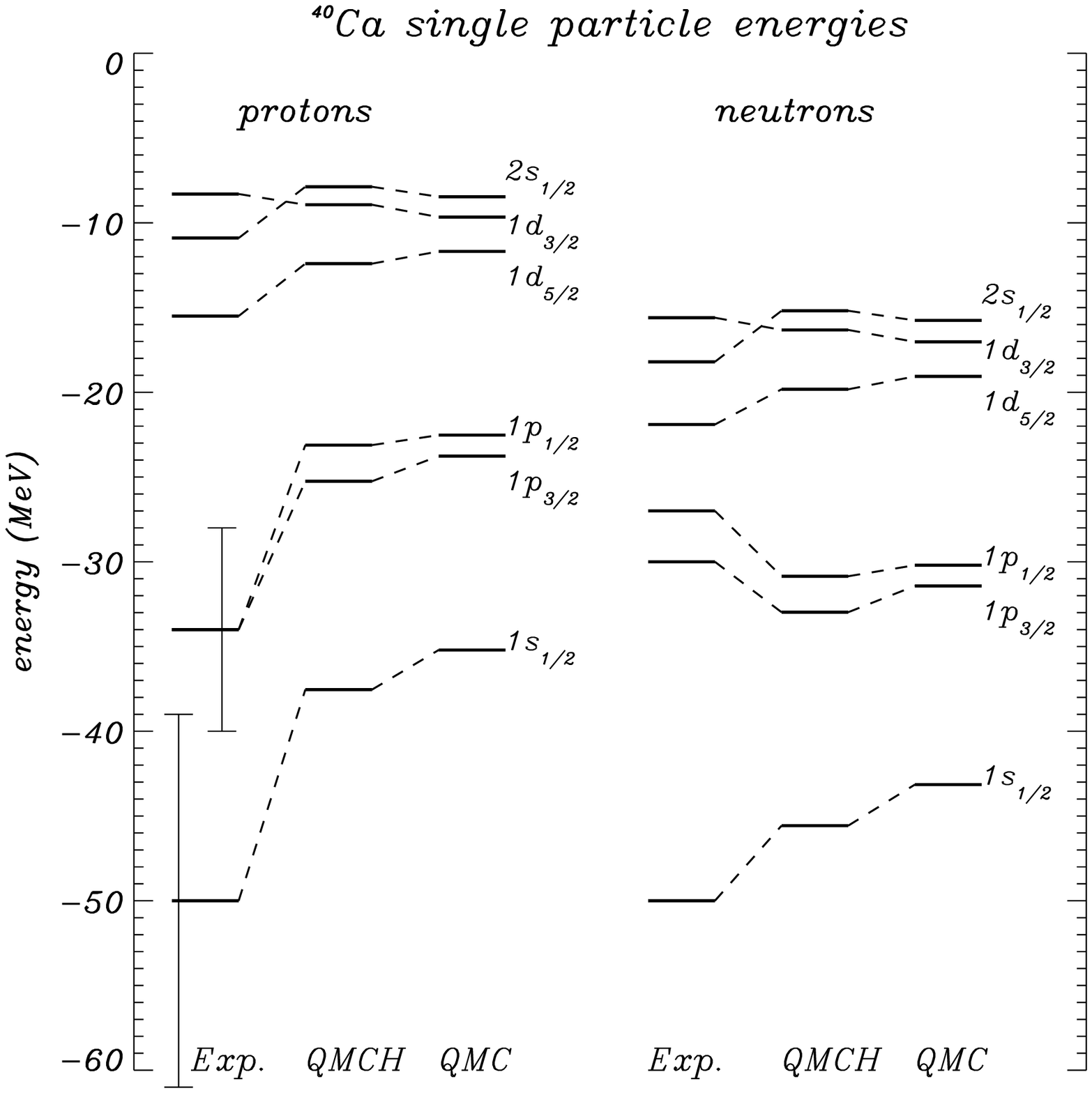,height=13cm}
\caption{Model predictions for the energy spectrum for $^{40}$Ca. 
QMC(H) denotes the case for $m_q$ = 5 (300) MeV and $R_B$ = 0.8 fm. }
\label{casp2}}
\end{figure}

\begin{figure}[htb]
\centering{\
\epsfig{file=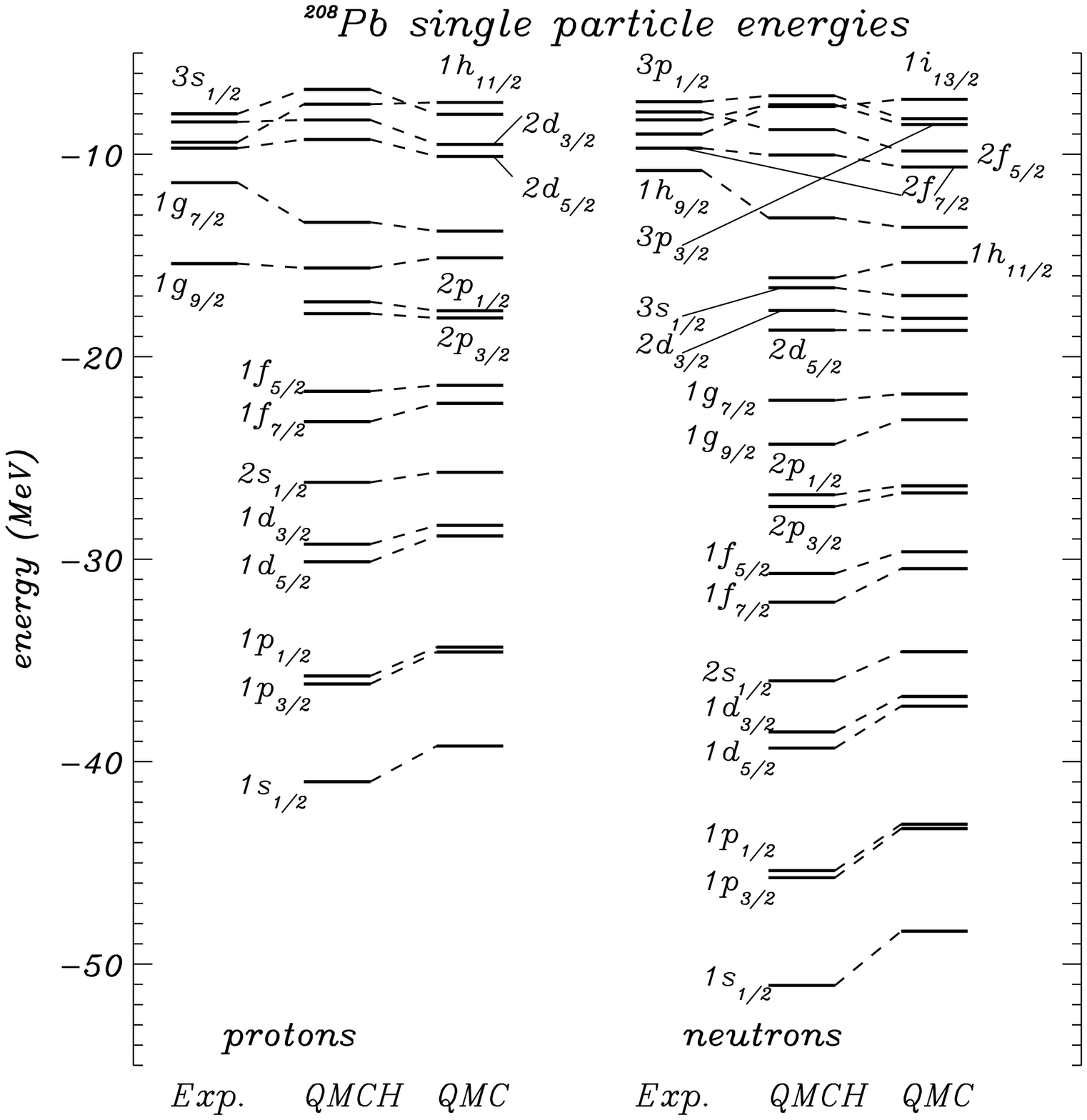,height=13cm}
\caption{Same as Fig.18 (for $^{208}$Pb). }
\label{pbsp2}}
\end{figure}

In Figs.\ref{casp2} and \ref{pbsp2}, the calculated spectra of 
$^{40}$Ca and $^{208}$Pb are presented.
Because of the relatively smaller scalar and vector fields in the present 
model than in QHD, the spin-orbit splittings are smaller.  
The improvement over QHD in the binding energy per nucleon comes 
at the expense of a reduction in the spin-orbit force.  
We should note that there is a strong correlation between the effective 
nucleon mass and the spin-orbit force.  This problem is also discussed 
in Refs.\cite{jin,blun}.  

As a test of the sensitivity of the spin-orbit splitting to features of
the model, 
we consider the case of a larger quark mass (noting that the Born-Oppenheimer 
approximation requires that the nucleon motion is 
relatively slow and the quarks are highly relativistic).  For example, 
we have calculated the case $m_q$ = 300 MeV (and $R_B$ = 0.8 fm) which 
is a typical constituent quark mass. 
The calculated spectra for $m_q$ = 300 MeV are also illustrated 
in Figs.\ref{casp2} and \ref{pbsp2} (QMCH).  
In this case, the various parameters were 
$g_\sigma^2/4\pi$ = 5.58, $g_\omega^2/4\pi$ = 8.51 (to 
satisfy the saturation condition), $g_\rho^2/4\pi$ = 6.45, 
$m_\sigma$ = 497 MeV (to fit the rms charge radius of $^{40}$Ca), 
$K$ = 334 MeV and $M_N^{\star}$ = 674 MeV at saturation density.  
We should record that the bag radius in this case 
increases by 7\% at saturation density (which is uncomfortably large).  
The slope parameter in Eq.(\ref{paramC}) is $a=3.9 \times 10^{-4}$.  
One can expect that a heavy quark mass gives a spectrum closer to 
those of QHD\cite{st1,finite}.  We can see from the figures 
that the calculated spectra are somewhat closer to the experimental data. 
We note that the charge density distributions for $^{40}$Ca and $^{208}$Pb 
are also reproduced well in this case.  

\clearpage

\section{Summary, discussion and further applications}
\label{discuss}
Starting with the quark-meson coupling model, 
in which quarks 
confined in nucleon bags interact through the exchange of scalar and 
vector mesons, we have presented a generalisation of QHD with a $\sigma$ 
field-dependent scalar coupling to describe finite nuclei quantitatively. 
The physical origin of this field-dependence, 
which provides a new saturation mechanism for nuclear matter, 
is the relatively rapid increase of the lower Dirac component of the 
wavefunction of the confined, light quark.  We have then derived a set of 
coupled non-linear differential equations which must be solved 
self-consistently but which are not much more difficult to solve than the 
relativistic Hartree equations of QHD.  Our calculations for static, 
closed-shell nuclei from $^{16}$O to $^{208}$Pb reproduce fairly well the 
observed charge density distributions, neutron density distributions etc.  

It will be very interesting to explore the connection between the $\sigma$ 
field-dependence of the variation of the effective $\sigma$-N 
coupling constant, which arises so naturally here, and the variation 
found empirically in earlier work. We note, in particular, that while 
our numerical results depend on the particular model chosen here (namely, 
the MIT bag model), the qualitative features which we find (such as the 
$\sigma$ field-dependent decrease of the scalar coupling etc.) will apply in 
{\em any model} in which the nucleon contains light quarks and the 
attractive N-N 
force is a Lorentz scalar.  Of course, it will be important to 
investigate the degree of variation in the numerical results for other 
models of nucleon structure\cite{blun}.

In the present model there are, however, still some discrepancies in energy 
spectra of nuclei, in particular, the spin-orbit splittings.  
In the previous section we have briefly discussed one possibility to 
partly overcome this defect in the present model.  
As an alternative approach, 
Jin and Jennings\cite{jin} and Blunden and Miller\cite{blun} have 
considered variations of the bag constant $B$ and $z$ parameter in 
nuclear matter, which have been suggested by the fact that quarks are 
presumably deconfined at high enough densities.  
This was taken to suggest that 
$B$ might decrease with increasing density.  Adopting this idea,  
Blunden and Miller\cite{blun} have studied properties of both nuclear 
matter and finite nuclei.  In their approach, $B$ in matter, $B^{\star}$, 
is given by 
\bge
B^{\star} = B \left[ 1 - \alpha_B \frac{U_s({\vec r})} 
{M_N} \right] , \label{B*}
\ene
where $U_s$ is an average scalar potential and $\alpha_B$ is an arbitrary 
parameter.  For finite nuclei the results move towards those of QHD, and 
there is an improvement in the spin-orbit splittings.  However, the 
nuclear incompressibility, $K$, is again larger than 
the experimental data, and the bag radius at saturation density increases 
by 5\% $\sim$ 13{\%}\cite{blun} or more\cite{jin}, which seems uncomfortably 
large\cite{emc}. 
We probably ought to construct a formalism 
beyond the Hartree approximation for this class of models in the future.  

In this work we have not considered any effect of a decrease of the meson 
masses in nuclear matter.  (In Ref.\cite{hadrons} these 
effects were investigated  
in infinite nuclear matter, using an earlier version of the
model.)  The $\sigma$-meson mass which we found it necessary to use in 
this paper ranges from  400 $\sim$ 450 MeV.  It seems rather small 
compared with 
the empirical $\sigma$-mass in the OBEP analyses of the free NN force 
\cite{formf,mach,obep}.  
On the other hand, noting its origin in two-pion-exchange with N-$\Delta$
intermediate states, we observe that the N-$\Delta$ mass difference
would decrease in medium by a similar percentage.
It will be interesting, in future work, to incorporate a self-consistent
reduction of the vector-meson mass \cite{hadrons} (as well 
as the scalar-meson mass\cite{sgmms}) 
into the calculation of the properties of finite nuclei.  

The successful generalisation of the QMC model to 
finite nuclei opens a tremendous number of opportunities for further 
work.  For example, earlier results for the Okamoto-Nolen-Schiffer 
anomaly\cite{st2}, the nuclear EMC effect\cite{st4}, 
the charge-symmetry violating correction to
super-allowed Fermi beta-decay\cite{st3}
and so on, can now be treated in a truly quantitative way. 
Since the present model can provide self-consistent quark wave functions in 
a deeply bound nucleon, e.g., $1s_{1/2}$ in $^{208}$Pb, it would also be very 
interesting to investigate variations in magnetic moments, $g_A$ and so on, 
for such nucleons\cite{yamazaki}. 

For hypernuclei the
natural extension (see Ref.\cite{hadrons}) is to assume that the 
$\sigma$ and $\omega$ mesons couple only to the non-strange 
constituents. From our discussion of the spin-orbit force in 
Ref.\cite{finite} and 
the fact that the spin of the $\Lambda$ is carried entirely by the 
strange quark, one can easily see that the $\Lambda$ spin-orbit force 
will arise entirely from the Thomas precession term.  
This means that the $\Lambda$ spin-orbit force is very naturally suppressed 
in this model -- as observed experimentally. It will be important to
follow this observation with quantitative results. 

Finally, we list a number of important ways in which this model could be
extended: replacing the
MIT bag by a model respecting PCAC (e.g., the 
cloudy bag model\cite{cbm}); replacing $\sigma$-exchange by 
two-pion exchange; replacing $\omega$ exchange by nucleon 
overlap at short distance, and so on. 
In terms of further theoretical development it will also be interesting to 
compare the present model with more phenomenological, non-linear 
extensions of QHD -- as reviewed recently in Ref.\cite{delf}.  
On the practical 
side, we stress that the present model can be applied to all the 
problems for which QHD has proven so attractive, with very little extra 
effort.

\vspace{1cm}
\flushleft{Acknowledgements} \\
\noindent
The authors are pleased to thank Pierre A.M. Guichon for valuable 
discussions and comments.  
This work was supported by the Australian Research Council. 
\end{document}